\documentclass[aps, prl, twocolumn, superscriptaddress,nobibnotes]{revtex4-1}
\usepackage{graphicx}
\usepackage[hidelinks]{hyperref} 
\usepackage{bm}
\usepackage{amsmath}
\usepackage{bbold}
\usepackage{xcolor}

\begin{document}

%\preprint{}

\title{Codimension-Two Spiral Spin-Liquid in the Effective Honeycomb-Lattice Compound Cs$_3$Fe$_2$Cl$_9$} 
\thanks{This manuscript has been authored by UT-Battelle, LLC under Contract No. DE-AC05-00OR22725 with the U.S. Department of Energy.  The United States Government retains and the publisher, by accepting the article for publication, acknowledges that the United States Government retains a non-exclusive, paid-up, irrevocable, world-wide license to publish or reproduce the published form of this manuscript, or allow others to do so, for United States Government purposes.  The Department of Energy will provide public access to these results of federally sponsored research in accordance with the DOE Public Access Plan (http://energy.gov/downloads/doe-public-access-plan).}

\renewcommand*{\thefootnote}{\arabic{footnote}}

\author{Shang Gao}
\email[]{sgao@ustc.edu.cn}
\affiliation{Department of Physics, University of Science and Technology of China, Hefei, Anhui 230026, People's Republic of China}
\affiliation{Materials Science \& Technology Division, Oak Ridge National Laboratory, Oak Ridge, TN 37831, USA}
\affiliation{Neutron Scattering Division, Oak Ridge National Laboratory, Oak Ridge, TN 37831, USA}

\author{Chris Pasco}
% \email[]{sgao.physics@gmail.com}
\affiliation{Materials Science \& Technology Division, Oak Ridge National Laboratory, Oak Ridge, TN 37831, USA}

\author{Otkur Omar}
\affiliation{Department of Physics, University of Science and Technology of China, Hefei, Anhui 230026, People's Republic of China}

\author{Qiang Zhang}
\affiliation{Neutron Scattering Division, Oak Ridge National Laboratory, Oak Ridge, TN 37831, USA}

\author{Daniel M. Pajerowski}
\affiliation{Neutron Scattering Division, Oak Ridge National Laboratory, Oak Ridge, TN 37831, USA}

\author{Feng Ye}
\affiliation{Neutron Scattering Division, Oak Ridge National Laboratory, Oak Ridge, TN 37831, USA}

\author{Matthias Frontzek}
\affiliation{Neutron Scattering Division, Oak Ridge National Laboratory, Oak Ridge, TN 37831, USA}

\author{Andrew F. May}
%\email[]{sgao.physics@gmail.com}
\affiliation{Materials Science \& Technology Division, Oak Ridge National Laboratory, Oak Ridge, TN 37831, USA}

\author{Matthew B. Stone}
\email[]{stonemb@ornl.gov}
\affiliation{Neutron Scattering Division, Oak Ridge National Laboratory, Oak Ridge, TN 37831, USA}

\author{Andrew D. Christianson}
\email[]{christiansad@ornl.gov}
\affiliation{Materials Science \& Technology Division, Oak Ridge National Laboratory, Oak Ridge, TN 37831, USA}

\date{\today}

%\keywords{}

\begin{abstract}
A codimension-two spiral spin-liquid is a correlated paramagnetic state with one-dimensional ground state degeneracy hosted within a three-dimensional lattice. Here, via neutron scattering experiments and numerical simulations, we establish the existence of a codimension-two spiral spin-liquid in the effective honeycomb-lattice compound Cs$_3$Fe$_2$Cl$_9$, which demonstrates a novel path to spiral spin-liquids by overcoming the long-standing impediment of weak further-neighbor interactions. In the long-range ordered regime, competing spiral and spin density wave orders emerge as a function of applied magnetic field, among which a possible order-by-disorder transition is identified.
\end{abstract}
\maketitle

\textit{Introduction.} 
A spiral spin-liquid (SSL) is an exotic type of correlated paramagnetic state where the low energy dynamics consist of collective spiral correlations~\cite{bergman_order_2007, lee_theory_2008, mulder_spiral_2010, zhang_exotic_2013, niggemann_classical_2019, attig_classical_2017, balla_degenerate_2020, balla_affine_2019, lee_theory_2008, chen_quantum_2017, yao_generic_2021, huang_versatility_2021, buessen_quantum_2018, liu_featureless_2020, mohylna_spontaneous_2022}. A characteristic feature of a SSL is that the  propagation vectors of the degenerate spiral ground states form a continuous surface in reciprocal space~\cite{bergman_order_2007}. Such an unusual yet clearly defined feature has stimulated a strong interest by the community to experimentally identify and understand SSLs in real materials~\cite{tristan_geometric_2005, suzuki_melting_2007, krimmel_spin_2009, matsuda_disordered_2010, macdougall_kinetically_2011, zaharko_spin_2011, nair_approaching_2014, macdougall_revisiting_2016, ge_spin_2017, graham_experimental_2023, gao_spiral_2017, gao_fractional_2020, rosales_aniso_2022, guratinder_magnetic_2022, chamorro_frustrated_2018, bai_magnetic_2019, tsurkan_on_2021, haraguchi_frustrated_2019, abdeldaim_realizing_2020, otsuka_canting_2020, wessler_observation_2020, bordelon_frustrated_2021, balz_physical_2016, pohle_theory_2021, takahashi_spiral_2024, gao_spiral_2021, cole_extreme_2023, gao_line_2022, hsieh_helical_2022, spiral_wan_2024, andriushin_observation_2025}. Through experimental and theoretical studies, compounds with a bipartite lattice, \textit{e.g.} the honeycomb~\cite{balz_physical_2016, pohle_theory_2021, gao_spiral_2021, cole_extreme_2023} and diamond~\cite{graham_experimental_2023, balz_physical_2016, pohle_theory_2021, gao_spiral_2017,bai_magnetic_2019, guratinder_magnetic_2022} lattices, have been demonstrated as the most fertile hosts of SSLs, where spiral correlations arise from the competition between the inter- and intra-sublattice interactions.

%---------------------------------------------------------------------
\begin{figure}[t!]
    \includegraphics[width=0.47\textwidth]{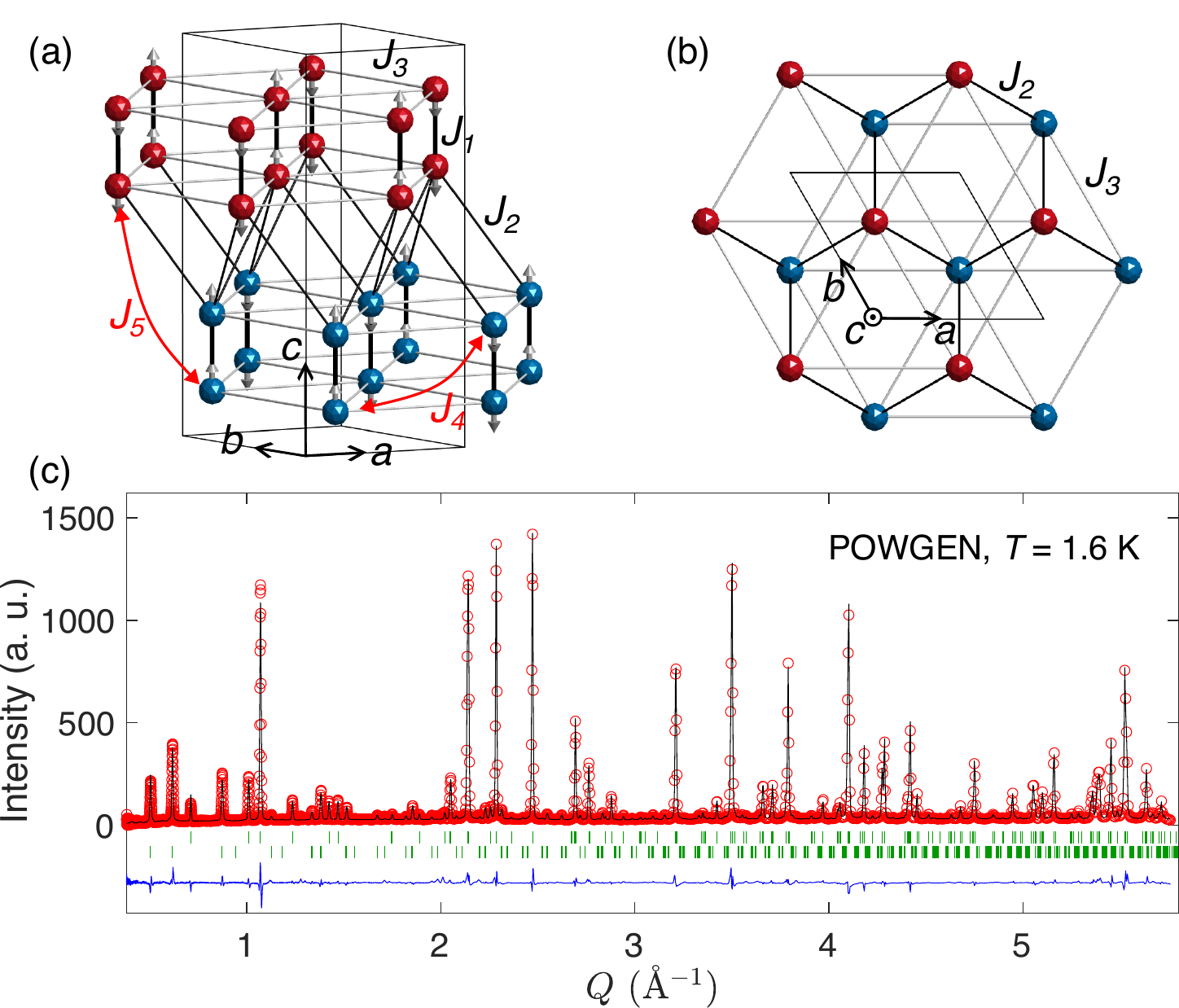}
    \caption{(a) AB-stacked triangular bilayers formed by the Fe$^{3+}$ ions in Cs$_3$Fe$_2$Cl$_9$. Atoms belonging to the neighboring bilayers are shown in red and blue, respectively. The $J_1$, $J_2$, and $J_3$ bonds are shown by thick black lines, thin black lines, and thin gray lines, respectively. The $J_4$ and $J_5$ bonds are indicated by red curved lines. Arrows indicate the spin directions of the collinear ground state with $\bm{q} = (\frac{1}{2}, 0, 0)$. (b) The AB-stacked triangular bilayers viewed along the $c$ axis. (c) Refinement result of the powder neutron diffraction data measured on POWGEN at $T$ = 1.6~K. Data points are shown as red circles. The calculated pattern is shown as the black solid line. The vertical bars indicate the positions of the structural (upper) and magnetic (lower) Bragg peaks for Cs$_3$Fe$_2$Cl$_9$. The blue line at the bottom shows the difference of measured and calculated intensities. The goodness-of-fit parameters are $R_\textrm{p}=18.6\%$ and $R_\textrm{wp}=10.2\%$.
    \label{fig:order}}
\end{figure}
%---------------------------------------------------------------------

%---------------------------------------------------------------------
\begin{figure*}[t]
  \includegraphics[width=0.95\textwidth]{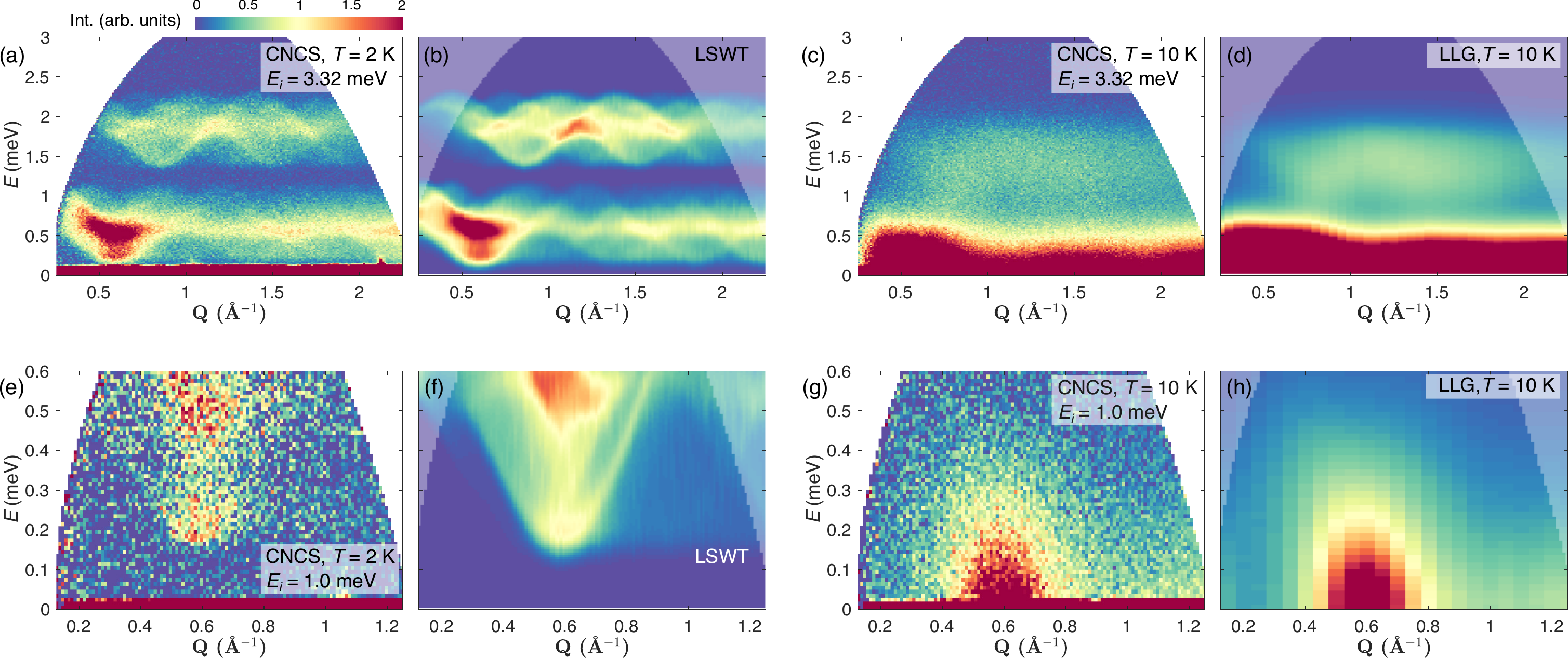}
  \caption{(a,c,e,g) Inelastic neutron scattering spectra $S(q,\omega)$ for Cs$_3$Fe$_2$Cl$_9$ powders measured on CNCS with $E_i=3.32$~meV at $T=2$ (a) and 10~K (c), and with $E_i=1.0$~meV at $T=2$ (e) and 10~K (g). (b,f) Simulated spectra for the fitted $J_{1\textrm{-}5}$-$D_z$ model using linear spin wave theory (LSWT). The spectra in (b) and (f) are convolved with the instrumental energy resolution for $E_i = 3.32$ and 1.0~meV, respectively. (d,h) Simulated spectra for the fitted $J_{1\textrm{-}5}$-$D_z$ model using the Landau-Langevin-Gilbert (LLG) method calculated at $T = 10$~K.
  \label{fig:cncs}}
\end{figure*}
%---------------------------------------------------------------------

Identifying novel SSL hosts is crucial for the realization of exotic spin textures like skyrmions~\cite{gao_fractional_2020, rosales_aniso_2022, takeda_magnon_2024} and subdimensional quasiparticles like fractons~\cite{yan_low_2021, pretko_fracton_2020}, and will also establish new candidate compounds to study the thermal and quantum order-by-disorder (ObD) transitions that are elusive in real materials~\cite{villain_order_1980, henley_order_1989, green_quantum_2018, bergman_order_2007, mulder_spiral_2010}. According to theoretical studies, SSLs can be classified by their codimension, a quantity that characterizes the dimensional difference between the spiral surface and the host system~\cite{yao_generic_2021}. Experimentally identified SSLs, including those observed on the diamond and honeycomb lattices, exhibit either a two-dimensional (2D) spiral surface on a three-dimensional (3D) lattice~\cite{gao_spiral_2017,bai_magnetic_2019,guratinder_magnetic_2022, gao_line_2022, graham_experimental_2023} or a one-dimensional (1D) spiral surface, i.e. a degenerate line, on a 2D lattice~\cite{balz_physical_2016,pohle_theory_2021,takahashi_spiral_2024,gao_spiral_2021}, thus all falling in the codimension one category. Although codimension-two SSLs have been predicted to exist on the AB-stacked
triangular lattice~\cite{yao_generic_2021, liu_classical_2016, hoang_hexagonal_2012, rastelli_degenerate_1983}, their experimental realizations  remain an open question. If proven true, this novel approach to spiral spin-liquids, where the intra-sublattice coupling arises from the relatively strong intralayer interactions, will relieve the long-standing obstacle of insufficient frustration on the original honeycomb lattice~\cite{gao_spiral_2021}.

Recent transport and magnetic characterizations of Cs$_3$Fe$_2$Cl$_9$ unveil possible honeycomb physics on a 3D lattice~\cite{ishii_field_2021}. In this compound, magnetic Fe$^{3+}$ ions with spin $S=\frac{5}{2}$ form AB-stacked triangular bilayers as shown in Figs.~\ref{fig:order}(a) and (b). Under dominant ferromagnetic (FM) interactions with exchange energy strength $J_1$ within the bilayers, a codimension-two SSL with a 1D spiral surface in the integer-$l$ planes may exist for $|J_3/J_2|>1/6$, where the threshold ratio is the same as that of the codimension-one SSL on the original honeycomb lattice~\cite{mulder_spiral_2010}. Although the magnetic ground state and spin dynamics in Cs$_3$Fe$_2$Cl$_9$ remain unexplored, a rich phase diagram has been established~\cite{ishii_field_2021}, suggesting the existence of strong magnetic frustration in this compound.

Here, through elastic and inelastic neutron scattering experiments on both single crystal and polycrystalline samples, we show that a codimension-two SSL with a uniaxial anisotropy is realized in Cs$_3$Fe$_2$Cl$_9$. The sprial surface's visibility is phase tuned as a function of the wavevector normal to the honeycomb plane. By combining neutron diffraction experiments and classical Monte Carlo simulations, we clarify the eight field-induced ordered phases as competing spiral and spin density wave (SDW) orders, among which a possible order-by-disorder transition is identified.

\textit{Magnetic ground state.} Powder neutron diffraction experiments were performed on POWGEN~\cite{huq_powgen_2011} at the Spallation Neutron Source (SNS) of the Oak Ridge National Laboratory (ORNL) to determine the magnetic LRO in Cs$_3$Fe$_2$Cl$_9$ below $T_N \sim 5.4$~K. Details on the sample preparation and neutron scattering experiments are presented in the Supplemental Material~\cite{supp}. As shown in Fig.~\ref{fig:order}(c), magnetic Bragg peaks belonging to the propagation vector $\bm{q}_\textrm{I}=(\frac{1}{2},0,0)$ are observed at low temperatures. Through Rietveld refinements~\cite{rodriguez_recent_1993}, the magnetic ground state is determined to be collinear as shown by arrows in Fig.~\ref{fig:order}(a), with an ordered moment size of 4.23(6)~$\mu_\textrm{B}$. This magnetic order is similar to the magnetic ground state of the isostructural Cs$_3$Fe$_2$Br$_9$~\cite{bruning_multiple_2021}.

\textit{Spin Dynamics and Modeling.} To determine the exchange coupling strengths, inelastic neutron scattering (INS) experiments were performed at CNCS~\cite{ehlers_new_2011s} at SNS of ORNL~\cite{supp}. Figure~\ref{fig:cncs}(a) presents the INS spectra collected with an incident neutron energy of $E_i=3.32$ (a,c) and 1.0~meV (e,g) at temperatures $T = 2$ (a,e) and 10~K (c,g), which are below and above $T_N$, respectively. Even for the powder sample, it is clear that there are two highly dispersive magnon modes centered around $E \sim 0.6$ and 1.8~meV energy transfer. From Fig.~\ref{fig:cncs}(e), an excitation gap of $\Delta \sim 0.2$~meV is observed at wavevector transfer $Q\sim0.6$~\AA$^{-1}$, suggesting the existence of a uniaxial single-ion anisotropy (SIA) that stabilizes the collinear ground state.

Using linear spin wave theory as implemented in the SpinW program~\cite{toth_linear_2015}, $\chi^2$-fits to the INS spectra were performed to analyze the spin interactions. As explained in the Supplemental Material~\cite{supp}, a minimal $J_{1\textrm{-}5}$-$D_z$ model with Hamiltonian $ \mathcal{H} = \sum_{\langle ij\rangle \in n}J_{n}\bm{S}_i\cdot\bm{S}_j + D_z(S_z)^2$ is employed in our calculations, which considers Heisenberg exchange interactions up to the fifth neighbors as shown in Fig.~\ref{fig:order} plus a uniaxial SIA term, $D_z$. As compared in Fig.~\ref{fig:cncs}(a,b,e,f), the $J_{1\textrm{-}5}$-$D_z$ model with fitted coupling strengths of $J_1 = -0.231(2)$, $J_2 = 0.082(2)$, $J_3 = 0.059(1)$, $J_4 = 0.0050(1)$, $J_5 = 0.0013(1)$ and $D_z = -0.032(2)$~meV reproduces the INS spectra. The dominant FM $J_1$ together with comparable strengths of $J_2$ and $J_3$ favor SSLs~\cite{mulder_spiral_2010,yao_generic_2021,liu_classical_2016}, while the relatively high magnitude of $|D_z|\sim0.54 J_3$ indicates the importance of the SIA in stabilizing the ground state.

The accuracy of the fitted model is further verified through the comparison of the INS spectra at $T = 10$~K shown in Fig.~\ref{fig:cncs}(c,d,g,h), where the calculations were performed using the Landau-Lifshitz method as implemented in the JuliaMD program~\cite{JuliaMD}. At elevated temperatures, the magnon mode at higher energy becomes broadened and softened, and the magnon band at lower energy collapses towards the elastic line, which indicates the emergence of strong spin correlation in the paramagnetic regime that can be directly detected through diffuse neutron scattering.

\begin{figure}[t!]
    \includegraphics[width=0.48\textwidth]{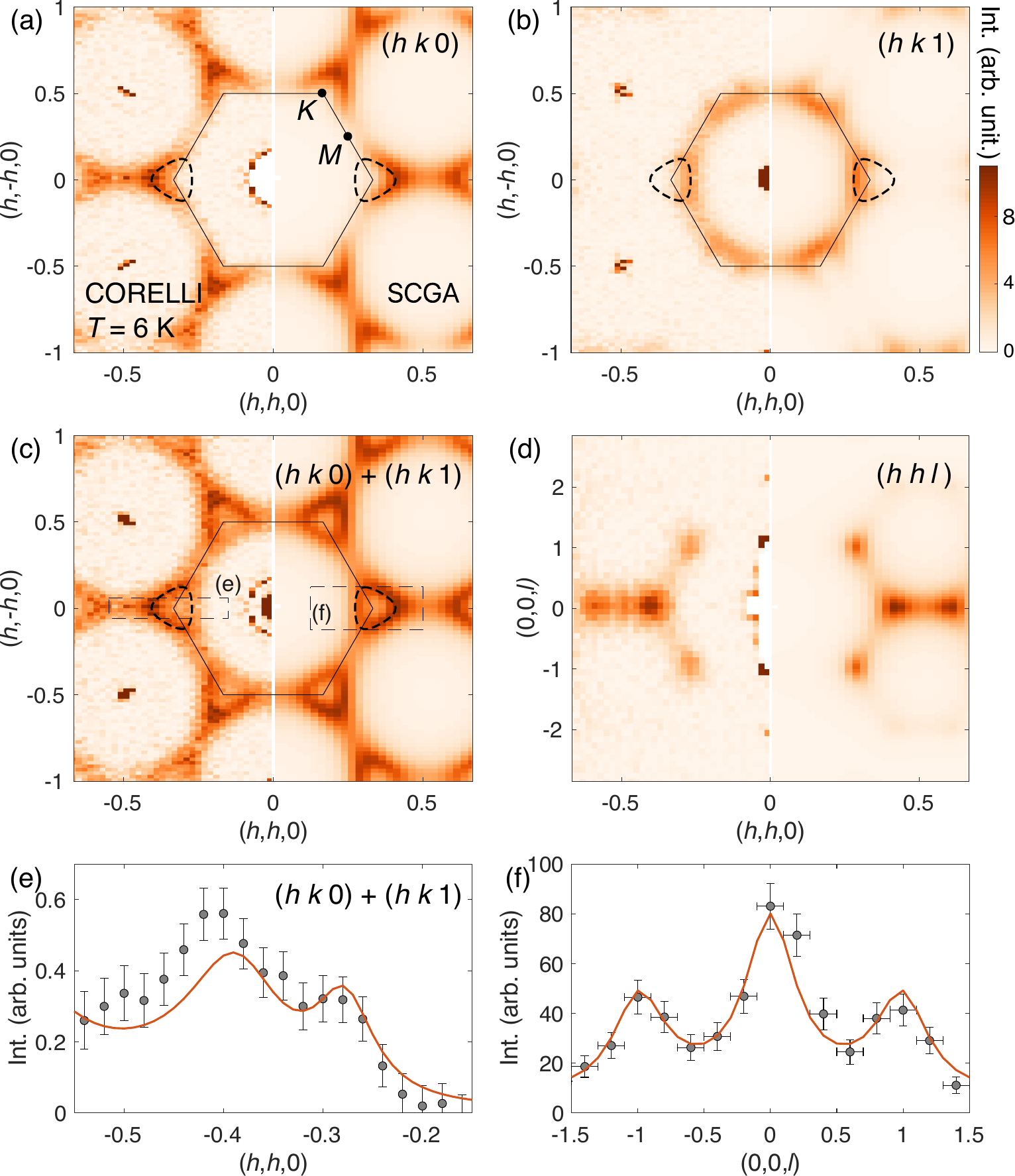}
    \caption{(a) Left half shows the diffuse neutron scattering pattern in the ($h$, $k$, 0) plane for Cs$_3$Fe$_2$Cl$_9$ measured at $T = 6$~K on CORELLI. Data are symmetrized according to the $6/mmm$ Laue class. The right panel is the calculated diffuse neutron scattering pattern for the fitted $J_{1\textrm{-}5}$-$D_z$ model using the SCGA method assuming a reduced $T = 5$~K to compensate for the underestimated critical correlations. Both the experimental and calculated data were integrated over $l=[-0.1,0.1]$ reciprocal lattice units, r.l.u. The solid-line hexagon indicates the boundary of the first Brillouin zone. Triangular-shaped lobes around the $K$-$(\frac{1}{3}, \frac{1}{3}, 0)$ points are the spiral surface for a honeycomb-lattice model with a frustration ratio of $J_2^\textrm{h}/J_1^\textrm{h} = 0.72$. (b) Similar experimental (left half) and calculated (right half) diffuse scattering patterns in the ($h$, $k$, 1) plane with an integration range of $l=[0.9,1.1]$~r.l.u. (c) Experimental (left half) and calculated (right half) diffuse scattering patterns with an integration range of $l=[-0.1,0.1]+[0.9,1.1]$~r.l.u. (d) Experimental (left half) and calculated (right half) diffuse scattering patterns in the $(h,h,l)$ plane with an integration width of 0.1 r.l.u. along $(h,-h,0)$. (e) Scattering intensity along $(h,h,0)$ integrated from the $(h,k,0)+(h,k,1)$ pattern in panel (c)  with an integration width of 0.06 r.l.u. along $(h,-h,0)$ as outlined by the dashed-line rectangle in the left half of panel (c). (f) $l$-dependence of the scattering intensity {integrated} in the area of $[0.125, 0.5]$ and $[-0.125, 0.125]$ r.l.u. along the $(h,h,0)$ and $(h,-h,0)$ directions, respectively. This area is outlined in the right half of panel (c) by a dashed-line rectangle. In panels (e) and (f), the red solid line shows the calculated results for the fitted $J_{1\textrm{-}5}$-$D_z$ model using the SCGA method.}
    \label{fig:corelli}
\end{figure}

\textit{Codimension-two SSLs.} Figure~\ref{fig:corelli}(a) presents the diffuse neutron scattering pattern for Cs$_3$Fe$_2$Cl$_9$ in the ($h$, $k$, 0) plane. Data were collected from CORELLI~\cite{ye_implementation_2018} at SNS of ORNL using an $\sim 8$~mg crystal~\cite{supp}. At $T = 6$~K, triangular shaped lobes are observed around the $K$-($\frac{1}{3}$, $\frac{1}{3}$, 0) points. As indicated by the dashed curves, the shape of the spiral surface is reproduced by a $J_1^h$-$J_2^h$ honeycomb-lattice model with $J_2^h/J_1^h=0.72$, where $J_1^h$ $(J_2^h)$ is equivalent to $J_2$ ($J_3$) in the $J_{1\textrm{-}5}$-$D_z$ model. The equal-time spin correlations for the fitted $J_{1\textrm{-}5}$-$D_z$ model can be calculated using the self-consistent Gaussian approximation (SCGA) method~\cite{canals_spin_2001, supp}. Since the critical correlations are underestimated in the SCGA method, a reduced $T = 5$~K is assumed in the calculations to better describe the experimental data. As compared in Fig.~\ref{fig:corelli}(a), the calculated and experimental intensity distributions agree well with each other, both following the spiral surface of a SSL on a honeycomb lattice.

%---------------------------------------------------------------------
%---------------------------------------------------------------------

%---------------------------------------------------------------------
\begin{figure*}[t]
    \includegraphics[width=0.9\textwidth]{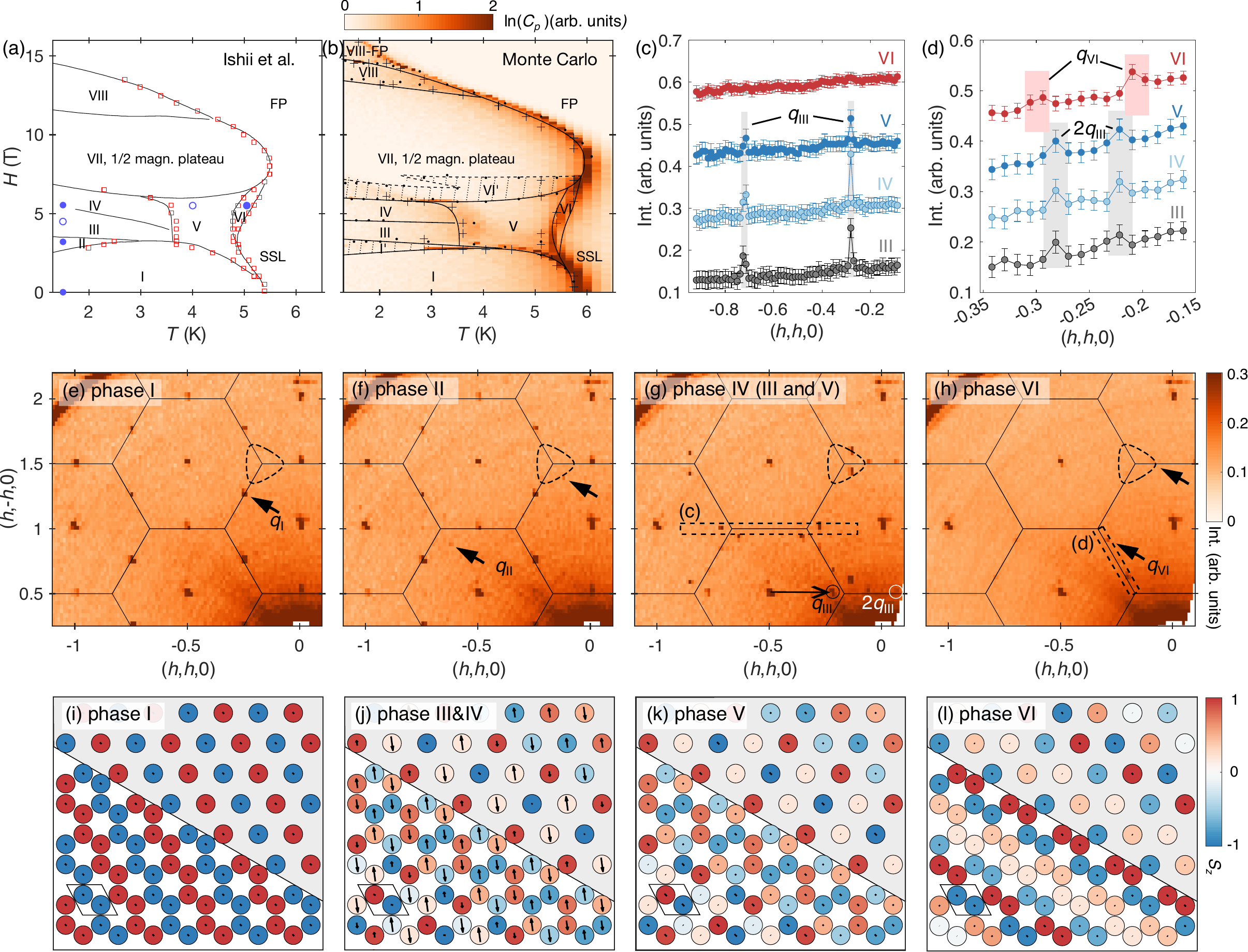}
    \caption{(a) $H$-$T$ phase diagram for Cs$_3$Fe$_2$Cl$_9$ reproduced from Ref.~\cite{ishii_field_2021} with magnetic field applied along the $c$ axis. Red squares are phase transitions revealed in heat capacity measurements~\cite{ishii_field_2021}. Blue points correspond to data described in the current work. Empty blue circles are the experimental conditions for measurements in phases III and V~\cite{supp}.  (b) $H$-$T$ phase diagram for the $J_{1\textrm{-}5}$-$D_z$ model obtained from classical Monte Carlo simulations. Pseudocolor corresponds to the calculated heat capacity $C_p$. Cross (Dot) marks are phase boundaries determined from $C_p(T)$ [$M(T)$]. (c) Diffraction intensity along the $(h+1,h-1,0)$ line as indicated by the dashed-line rectangle in panel (g) in phases III-VI. The integration width perpendicular to the line is 0.05~r.l.u. (d) Diffraction intensity along the $(-2h,4h,0)$ line as indicated by the dashed-line rectangle in panel (h) in phases III-VI. The integration width perpendicular to the line is 0.05~r.l.u. (e-h) Diffraction pattern in the ($h$, $k$, 0) plane collected on WAND$^2$ in (e) phase I with $T = 1.5$~K and $H = 0$~T, (f) phase II with $T = 1.5$~K and $H = 3.2$~T, (g) phase IV with $T = 1.5$~K and $H = 5.5$~T, and (h) phase VI with $T = 5.05$~K and $H = 5.5$~T. In each panel, reflections belonging to the characteristic propagation vectors are indicated by black arrows. In panel (g), the 2$\bm{q}$ reflection is indicated by the white circle. The shape of spiral surface is shown in dashed line. (i-l) Magnetic structures for phase (i) I, (j) III and IV, (k) V, and (l) VI. The $xy$ components of the ordered spins are indicated by black arrows. The $z$ components are encoded by colors. In each panel, the bottom left (top right) part depicts the spin configuration for two sublattices (one sublattice).
    \label{fig:wand}}
\end{figure*}
%---------------------------------------------------------------------

A special feature of the SSL in Cs$_3$Fe$_2$Cl$_9$ is the visibility of the spiral surface, which arises from its unique codimension two. On the original honeycomb lattice with antiferromagnetic $J_1^h$, the spiral surface within the first Brillouin zone has a structure factor of zero due to the interference between the two sublattices~\cite{liu_classical_2016}, leading to a diffuse pattern similar that shown in Fig.~\ref{fig:corelli}(a). However, as compared in Figs.~\ref{fig:corelli}(a) and (b), the visibility of the spiral surface in Cs$_3$Fe$_2$Cl$_9$ is complementary between the $l=0$ and $l=1$ planes. This variance in visibility originates from the $l$-dependence of the phase factor, which modulates the interference between the two sublattices and thus reflects the higher codimension of the SSL~\cite{liu_classical_2016}. By summing up the intensity in the planes of $l = 0$ and 1, the complete spiral surface can be recovered in Fig.~\ref{fig:corelli}(c), upon which a line cut along the ($h,~h,~0$) direction is shown in Fig.~\ref{fig:corelli}(e).

The scattering pattern in the ($h,~h,~l$) plane presented in Fig.~\ref{fig:corelli}(d) reveals weak scattering intensity between the integer-$l$ planes. According to our calculations, the intensity of the interplanar scattering diminishes with decreasing temperatures, thus can be attributed to thermal excitations out of the ground state manifolds.  This observation is further confirmed in Fig.~\ref{fig:corelli}(f) through the measured and calculated $l$-dependence of the integrated intensity within a rectangular area of $[0.125, 0.5]$ r.l.u. and $[-0.125, 0.125]$ r.l.u. along the $(h,h,0)$ and $(h,-h,0)$ directions, respectively.

\textit{Competing orders and a possible ObD transition.} In the absence of SIA, SSLs on a honeycomb lattice have been predicted to exhibit an ObD transition at low temperatures~\cite{bergman_order_2007, mulder_spiral_2010, liu_featureless_2020}. In the regime of $J_2^h/J_1^h>0.5$, the magnetic propagation vector $\bm{q}$ of the ObD phase lies at the corners of the triangular-shaped spiral surface~\cite{ mulder_spiral_2010, liu_featureless_2020}. This is obviously not the case for Cs$_3$Fe$_2$Cl$_9$ with $\bm{q}_\textrm{I} = (\frac{1}{2}, 0, 0)$ as revealed in Fig.~\ref{fig:order}(c), suggesting the strong impact of the SIA. 

However, the rich phase diagram reported for Cs$_3$Fe$_2$Cl$_9$~\cite{ishii_field_2021}, which is also reproduced in Fig.~\ref{fig:wand}(a), indicates the the possibility of an ObD in magnetic field. In fields of $3 \lesssim H \lesssim 6$~T, multiple phases, II-VI, emerge. Two additional phases exist at $H\gtrsim 6$~T, including a $\frac{1}{2}$-magnetization plateau phase, VII, up to $\sim 12$~T and a transitional phase, VIII, that precedes the field-polarized FM phase at $H\gtrsim15$~T.

Single crystal neutron diffraction experiments were performed on WAND$^2$~\cite{frontzek_wand_2018} at the High Flux Isotope Reactor (HFIR) at ORNL to clarify the LRO phases as a function of magnetic field up to 6~T~\cite{supp}. As summarized in Figs.~\ref{fig:wand}(e)-\ref{fig:wand}(h), four different types of diffraction patterns are observed in the ($h$, $k$, 0) plane among phases I-VI. Compared to that in phase I, the diffraction pattern of phase II in Fig.~\ref{fig:wand}(f) exhibits additional weak magnetic Bragg peaks at $\bm{q}_\textrm{II}=(\frac{1}{4},\frac{1}{4},0)$, indicating the coexistence of minor $\uparrow\uparrow\uparrow\downarrow$ and major $\uparrow\downarrow\uparrow\downarrow$ magnetic domains, where $\uparrow$ ($\downarrow$) represents spins (antiparallel) parallel with $c$. Surprisingly, the diffraction patterns in phases III-V, among which the pattern for phase IV is shown in Fig.~\ref{fig:wand}(g), are similar to each other~\cite{supp}: all exhibiting main reflections at $\bm{q}_\textrm{III} = \frac{5}{6}\times(\frac{1}{3}, \frac{1}{3}, 0)$ close to the center of the edges of the triangular-shaped spiral surface as indicated by dashed lines, together with weaker secondary reflections at $2\bm{q}_\textrm{III}$. The existence of the latter often suggests a field-distorted spiral or SDW phase~\cite{gignoux_neutron_1998, stusser_neutron_2002}.

As shown in Fig.~\ref{fig:wand}(h) and also compared in Fig.~\ref{fig:wand}(c) for the line cuts along the $(h+1,h-1,0)$ direction, the $\bm{q}_\textrm{III}$ refelctions disappear in phase VI. Instead, new magnetic reflections belonging to $\bm{q}_\textrm{VI} = \frac{11}{9}\times(\frac{1}{3}, \frac{1}{3}, 0)$ appear at the corners of the triangular-shaped spiral surface, which is exactly the position predicted by the ObD transition~\cite{mulder_spiral_2010, liu_featureless_2020}. Comparison of the line cuts along the $(-2h,4h,0)$ direction shown in Fig.~\ref{fig:wand}(d) reveal a clear shift in the $\bm{q}_\textrm{VI}$ and $2\bm{q}_\textrm{III}$ positions, evidencing the distinct magnetic order in phase VI.

Using classical Monte Carlo simulations in a parallel tempering scheme as implemented in the SpinMC program~\cite{SpinMC}, we calculated the $H$-$T$ phase diagram for the fitted $J_{1\textrm{-}5}$-$D_z$ model. The calculated phase diagram, shown in Fig.~\ref{fig:wand}(b), agrees well with the experimental results shown in Fig.~\ref{fig:wand}(a), while some discrepancies existing in the transitional regime, including phases I' and VI' shadowed by dashed lines, may originate from the absence of quantum fluctuation in our calculations~\cite{supp}. The agreement between the experimental and calculated phase diagrams allows us to use the simulated magnetic structures as constraints in refining the magnetic orders against the neutron diffraction dataset collected in each phase~\cite{supp}.

Following this method, the magnetic orders in phases I-VI can be determined, and the representative structures in the neighboring bilayers are shown in Figs.~\ref{fig:wand}(i)-(l). Phases III and IV are determined to be of an elongated spiral-type orders, and the transition between them can be attributed to the modulation of the $\bm{q} = 0$ magnitude~\cite{supp}. As a contrast, both phases V and VI are revealed to be of the SDW orders with spins along the $c$ axis despite their different $\bm{q}$ vectors. Since $\bm{q}_\textrm{VI}$ in phase VI is the predicted propagation vector of the ObD theory~\cite{mulder_spiral_2010, liu_featureless_2020}, it is tantalizing to ascribe the transition from the SSL state to phase VI as entropy-driven. The proposed SDW character of phase VI, however, poses a challenge in accurately determining the free energy for states above $T_N$ under the current theoretical framework~\cite{bergman_order_2007}. Further studies will be required to clarify the character of the SSL-SDW phase transition.

\textit{Discussion}. From the perspective of experimental realization, the introduction of codimension into SSLs is not a mere classification for convenience but rather has important implications~\cite{yao_generic_2021}. On the original honeycomb lattice, the realization of SSL requires a relatively strong second-neighbor interactions, which is rarely satisfied in real materials~\cite{gao_spiral_2021}. However, as demonstrated in our work for Cs$_3$Fe$_2$Cl$_9$, the relatively strong intralayer couplings on the AB-stacked triangular lattice, which correspond to the second-neighbor interactions on an effective honeycomb lattice, naturally favor the emergence of SSLs. Besides the many compounds that are isostructural with Cs$_3$Fe$_2$Cl$_9$~\cite{bruning_multiple_2021, leuenberger_synthesis_1986, leuenberger_synthesis_1986_2, leuenberger_exchange_1986}, there are a rich variety of candidate systems like $R$ZnPO~\cite{spiral_wan_2024, nientiedt_equiatomic_1998} and $R_3$Ru$_4$Al$_{12}$~\cite{hirschberger_skyrmion_2019, gao_ordering_2019, kolincio_large_2021} where the SSL physics might be achieved. 

Codimension-two SSLs also offer a plethora of novel physics that remain to be explored. One immediate question is the impacts of thermal and quantum fluctuations over the SSL state. Besides the thermally induced ObD transitions, there are theoretical discussions about the prospect of realization quantum spin liquid on the honeycomb lattice through SSLs~\cite{reuther_magnetic_2011, niggemann_classical_2019}. On the AB-stacked triangular lattice, this approach may still be valid, which resonates with the recent efforts of utilizing the further-neighbor interactions on a triangular lattice to achieve quantum spin liquids near the phase boundaries~\cite{gallegos_phase_2025}. Another exciting direction to explore is the novel spin textures that may emerge from the codimension-two SSLs. On an AB-stacked triangular lattice, the inter- and intra-layer interactions may follow different symmetry constraints depending on the presence or location of the centrosymmetric points. This additional freedom in symmetry constraints can be highly beneficial for the realization of exotic spin textures like magnetic skyrmions~\cite{mohylna_spontaneous_2022, mohylna_frustration_2025}.

\begin{acknowledgments}
We acknowledge helpful discussions with Gang Chen, Xu-Ping Yao, and James Jun He. This work was supported by the U.S. Department of Energy, Office of Science, Basic Energy Sciences, Materials Sciences and Engineering Division. This research used resources at the Spallation Neutron Source (SNS) and the High Flux Isotope Reactor (HFIR), both are DOE Office of Science User Facilities operated by the Oak Ridge National Laboratory (ORNL). The proposal numbers are IPTS-28988 (POWGEN), 29069 (CNCS), 31251 (CORELLI), and 29937 (WAND$^2$). Numerical calculations in this paper were partly completed on the supercomputing system in the Supercomputing Center of USTC. Works at USTC were funded by the National Science Foundation of China (NSFC) under the Grant No. 12374152.
\end{acknowledgments}

% \bibliography{CFC_honeycomb}
%


\clearpage
\newpage

\renewcommand{\thefigure}{S\arabic{figure}}
\renewcommand{\thetable}{S\arabic{table}}
\renewcommand{\theequation}{S\arabic{equation}}

\makeatletter
\renewcommand*{\citenumfont}[1]{S#1}
\renewcommand*{\bibnumfmt}[1]{[S#1]}
\def\clearfmfn{\let\@FMN@list\@empty}  
\makeatother
\clearfmfn

\setcounter{figure}{0} 
\setcounter{table}{0}
\setcounter{equation}{0} 
\setcounter{section}{0} 

\onecolumngrid
\begin{center} {\bf \large Supplemental Materials for:\\Codimension-Two Spiral Spin-Liquid in the Effective Honeycomb-Lattice Compound Cs$_3$Fe$_2$Cl$_9$} \end{center}
  %\maketitle

\vspace{0.5cm}

\renewcommand*{\thefootnote}{\arabic{footnote}}
\renewcommand{\thefigure}{S\arabic{figure}}
\renewcommand{\thetable}{S\arabic{table}}
\renewcommand*{\thefootnote}{\arabic{footnote}}

\section{Powder sample synthesis}

Cs$_3$Fe$_2$Cl$_9$ is extremely air sensitive and thus all handling of reagents and products was performed inside an inert atmosphere glovebox.  Anhydrous FeCl$_3$ (as-received) and CsCl (dried in air at 400~°C) were loaded into a fused silica crucible.  The crucible was placed inside a fused silica ampoule, transferred to a vacuum line without exposure to air, purged with argon and sealed with approximately $\frac{1}{4}$ atm argon.  The reagents were heated to 700~°C at a rate of 60~°C/h and held for 8~h prior to quenching in an ice-water bath.  A boule was easily extracted from the silica ampoule, which was then ground and the powder was sealed with argon gas in a manner as done for the initial reaction.  The powder was annealed at 250~°C for 10~d, with an external thermocouple utilized to verify the annealing temperature; the furnace was turned off and the sample was allowed to cool naturally.

\section{Single crystal growth}

Single crystals of Cs$_3$Fe$_2$Cl$_9$ were synthesized through a solvothermal route with a variety of concentrations and temperature profiles tested.  One consistent feature of the growths was that a small excess of CsCl was found to improve the size and apparent quality of the crystals obtained.  In the best growth conditions 1 mmol of anhydrous FeCl$_3$ and 2 mmol anhydrous CsCl were weighed out in a glovebox under argon and removed in a small, closed vial.  This vial was then emptied into a 23 mL Parr A280AC PTFE liner in air with concentrated HCl added immediately after, with additional HCl being used as a wash of the vial to ensure all material was transferred, with a total of 5 mL concentrated HCl being added to PTFE liner, which was subsequently sealed in a Parr 4749 general purpose acid digestion vessel.  The vessel was placed in an oven which was ramped to 220~°C over one hour and held at this temperature for 6 hours to allow the vessel to come to temperature and allow the reactants to be fully dissolved in the concentrated HCl.   The vessel was then cooled to 30~°C over 48 hours before the oven was turned off, with slower cooling rates not providing significant improvements in crystal size.

Due to the pressures involved and the porosity of PTFE the exterior of the liner was coated in a water-soluble green chloride salt due to the corrosion of the acid digestion vessel which was removed before the PTFE liners were opened.  The liners were opened in air, with the resulting crystals metastable in the mother liquor, though undergoing a process of dissolution and recrystallization if stored under the mother liquor long-term.  To avoid this, and because the crystals were highly air-sensitive they were immediately transferred into concentrated HCl under which they could be safely stored and additionally helped to remove the mother liquor from the crystals.  The crystals stored under concentrated HCl were transferred to a glove bag which was filled with argon to create an air free environment safe from corrosion where the crystals were removed from the concentrated HCl, dried via filtration and transferred into a vial under argon which, free of HCl, was then transferred into a helium filled glovebox for long term storage.  Crystals obtained through this method were up to 25 mg, though the largest unambiguously single crystal was 7.5 mg which was selected for neutron diffraction experiments on CORELLI and WAND$^2$.

\section{Magnetization measurements}

Magnetization measurements were performed on a single crystal weighing 4.5(1) mg using a Quantum Design Magnetic Property Measurement System MPMS-3.  Due to the air sensitivity of the crystal it was coated in and secured to a quartz post with silicone vacuum grease.  Magnetization measurements revealed similar phase transitions as previously reported~\cite{ishii_field_2021s} and were used to plan the neutron diffraction experiment.

%---------------------------------------------------------------------
\begin{figure}[h!]
    \includegraphics[width=0.75\textwidth]{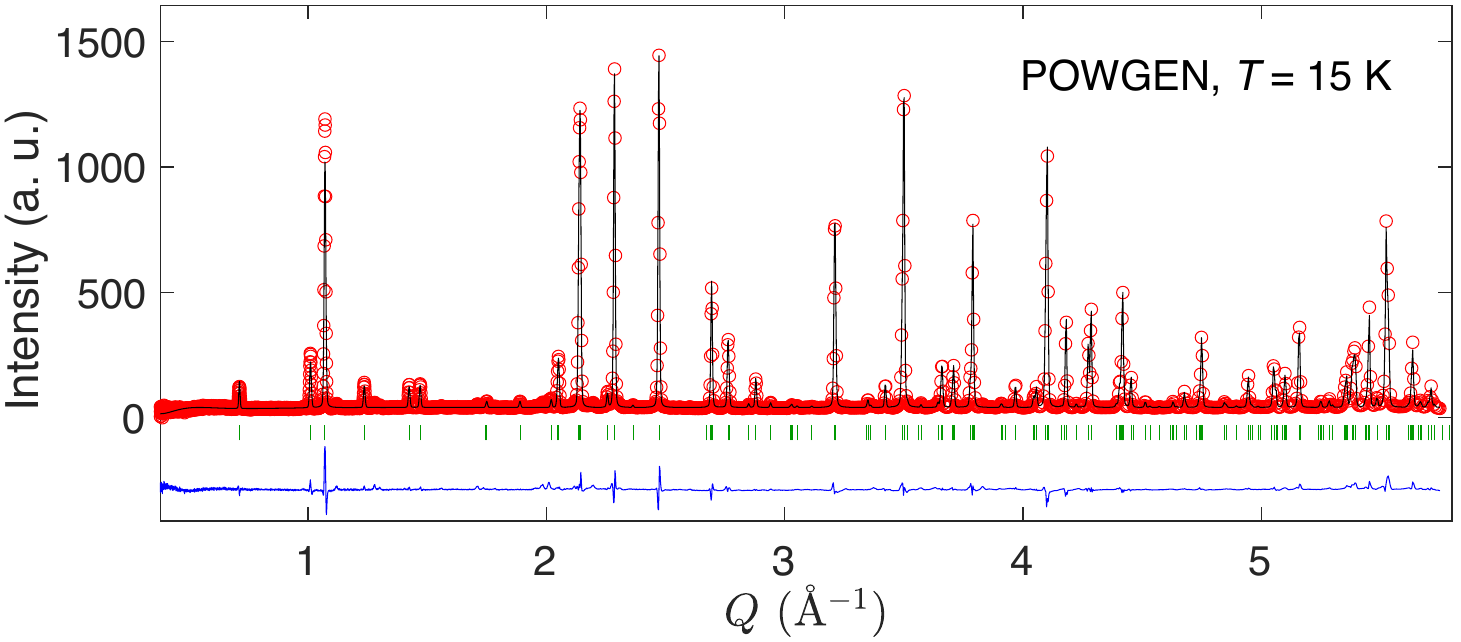}
    \caption{Refinement result of the powder neutron diffraction data measured on POWGEN at $T$ = 15~K. Data points are shown as red circles. The calculated pattern is shown as the black solid line. The vertical bars indicate the positions of the structural Bragg peaks for Cs$_3$Fe$_2$Cl$_9$. The blue line at the bottom shows the difference between measured and calculated intensities.
    \label{fig:refine}}
\end{figure}
%---------------------------------------------------------------------
%---------------------------------------------------------------------
\begin{figure}[h!]
    \includegraphics[width=0.75\textwidth]{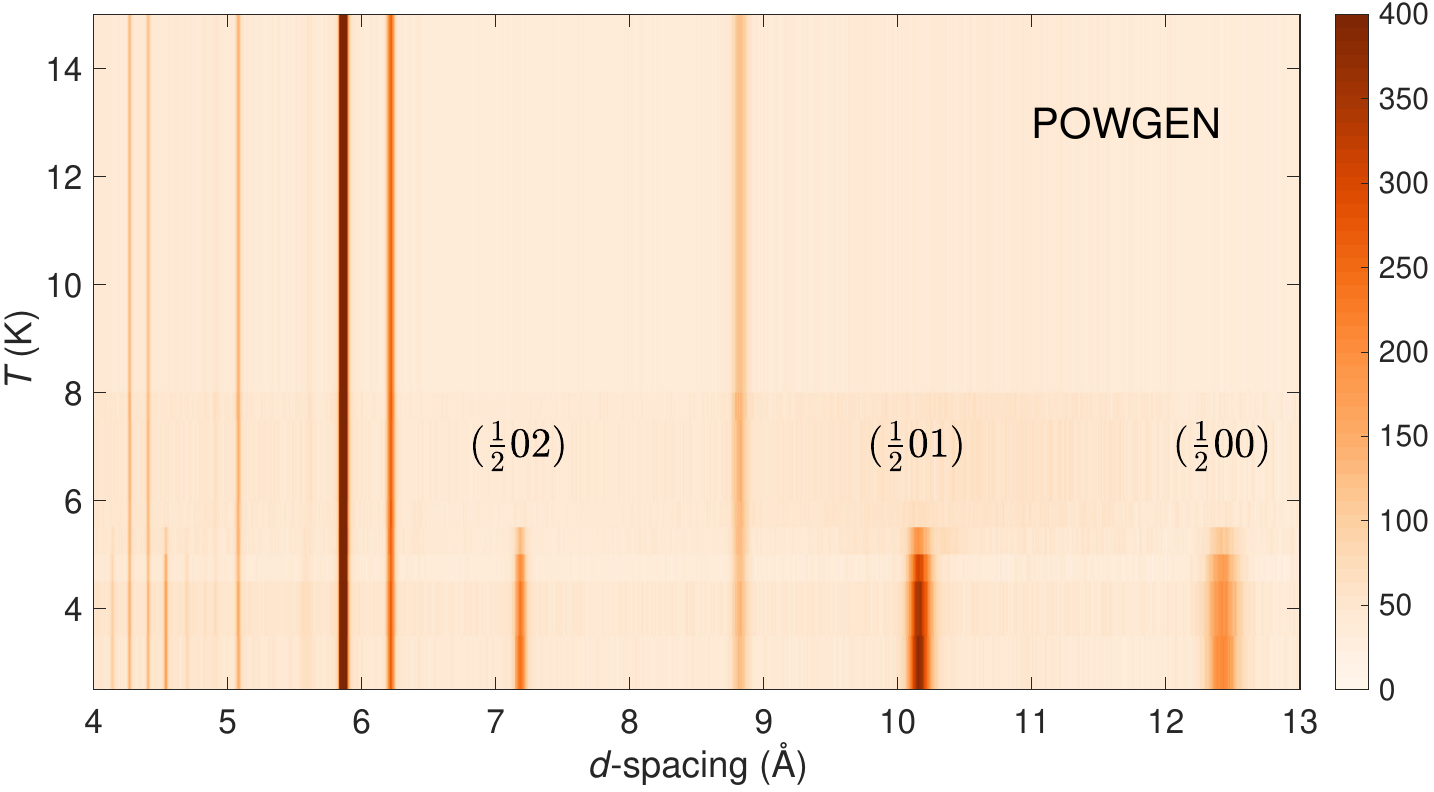}
    \caption{Temperature evolution of the powder neutron diffraction data for Cs$_3$Fe$_2$Cl$_9$ collected on POWGEN. Positions of the representative magnetic Bragg peaks $(\frac{1}{2}00)$, $(\frac{1}{2}01)$, and $(\frac{1}{2}02)$ are indicated.
    \label{fig:pg3}}
\end{figure}
%---------------------------------------------------------------------
\section{Neutron diffraction experiments on POWGEN}

Powder neutron diffraction experiments were performed on POWGEN~\cite{huq_powgen_2019} at the Spallation Neutron Source (SNS) of the Oak Ridge National Laboratory (ORNL). About 5~g powder of Cs$_3$Fe$_2$Cl$_9$ was filled into an air-tight vanadium can in a helium filled glovebox. An orange cryostat was utilized to reach a base temperature of 2~K. Data reduction was performed using the MANTID software~\cite{arnold_mantid_2014}. 

Figure~\ref{fig:refine} summarizes the refinement result of the neutron diffraction pattern collected at $T=15$~K. No secondary reflections are observed in the diffraction pattern, which confirms the phase purity of our sample. Refined crystal structure parameters are listed in Table~\ref{tab:refine}. Due to the lacking of neutron diffraction data in the high-$Q$ region, we cannot reliably fit the thermal parameters. Therefore, a uniform thermal factor of $B_\textrm{iso}= 0.1$~\AA$^2$ is assumed for all atoms.  Figure~\ref{fig:pg3} plots the temperature evolution of the diffraction pattern. Below $T_\textrm{N}\sim5.5$~K, magnetic Bragg peaks belonging to $\bm{q}=(\frac{1}{2},0,0)$ are observed. This transition temperature is consistent with that observed in magnetic susceptibility.

The refinement result of the neutron diffraction pattern collected at $T=1.6$~K is shown in Fig.~1 of the main text. Table~\ref{tab:refine2} summarizes the refined crystal structure parameters.  The refined magnitude of the ordered moment is 4.23(6)~$\mu_\textrm{B}$.

\begin{table}[h]
\caption{\label{tab:refine} Refined crystal structure parameters for Cs$_3$Fe$_2$Cl$_9$ at 15~K. The space group is $P6_3/mmc$ with lattice constants $a=7.1784(1)$ and $c=17.6374(3)$~\AA. A uniform thermal parameter of $B_\textrm{iso}= 0.1$~\AA$^2$ is assumed for all atoms. The goodness-of-fit parameters are $R_p=18.6\%$ and $R_{wp}=10.2\%$.\\}
\begin{tabular}{c c c c c}
\hline 
Atom &  $x$ &  $y$ &  $z$       & site  \\ \hline
Cs1  &  0   &   0  &   1/4      & $2b$  \\ 
Cs2  &  1/3 &  2/3 &  0.0826(2) & $4f$  \\ 
Fe   &  1/3 &  2/3 &  0.8458(1) & $4f$  \\ 
Cl1  &  0.519(4)   &  0.0330(3) &  1/4       & $12j$  \\ 
Cl2  &  0.8248(2)  & 0.6496(4)  &  0.0892(1) & $12k$  \\ 
\hline
\end{tabular}
\end{table}

\begin{table}[h!]
\caption{\label{tab:refine2} 
    Refined crystal structure parameters for Cs$_3$Fe$_2$Cl$_9$ at 1.6~K. The space group is $P6_3/mmc$ with lattice constants $a=7.1784(1)$ and $c=17.6369(2)$~\AA. A uniform thermal parameter of $B_\textrm{iso}= 0.1$~\AA$^2$ is assumed for all atoms. The goodness-of-fit parameters are $R_p=15.2\%$ and $R_{wp}=11.3\%$.\\}
\begin{tabular}{c c c c c}
\hline 
Atom &  $x$ &  $y$ &  $z$       & site  \\ \hline
Cs1  &  0   &   0  &   1/4      & $2b$  \\ 
Cs2  &  1/3 &  2/3 &  0.0829(2) & $4f$  \\ 
Fe   &  1/3 &  2/3 &  0.8459(1) & $4f$  \\ 
Cl1  &  0.510(4)   &  0.0325(3) &  1/4       & $12j$  \\ 
Cl2  &  0.8247(2)  & 0.6495(4)  &  0.0891(1) & $12k$  \\ 
\hline
\end{tabular}
\end{table}

\section{Inelastic neutron scattering experiments on CNCS}

Inelastic neutron scattering (INS) experiments on powder samples of Cs$_3$Fe$_2$Cl$_9$ were performed on CNCS~\cite{ehlers_new_2011} at the SNS of the ORNL. About 5~g powder was sealed in an aluminum can in a helium filled glovebox. An orange cryostat was utilized to reach a base temperature of 5~K. Measurements were taken with incident neutron energies of $E_i = 6.59$, 3.32, and 1.0~meV in the high flux chopper configuration at $T = 2$ and 10~K. For each measuring condition, data were collected on an empty can and subtracted as background. Data reduction and projection were performed using the Mslice program in DAVE~\cite{azuah_dave_2009}.

\section{Diffuse neutron scattering experiments on CORELLI}

Single crystal diffuse neutron scattering experiments were performed on CORELLI~\cite{ye_implementation_2018s} at the SNS of the ORNL. A crystal (mass $\sim7.5$\,mg) was aligned with the $c$ axis vertical. A closed cycle refrigerator (CCR) was employed to reach temperatures $T$ down to 5~K.  Data were acquired by rotating the sample in 1.5$^{\circ}$ steps, covering a total range of 360$^{\circ}$. The counting time at each rotation angle was approximately 1.5~mins with the source operating at a power of 1.4~MW. Data reduction and projection were performed using the MANTID software~\cite{arnold_mantid_2014}. Data collected at $T = 20$~K were subtracted as background. Theoretical diffuse neutron scattering patterns were calculated using the self-consistent Gaussian approximation (SCGA) method~\cite{conlon_absent_2010} as implemented in JuliaSCGA~\cite{JuliaSCGA, gao_spiral_2021s}.

\section{Neutron diffraction experiments on WAND$^2$}

Single crystal neutron diffraction measurements were performed using the WAND$^2$ diffractometer~\cite{frontzek_wand_2018s} at the High Flux Isotope Reactor HFIR with $ \lambda = 1.486$~\AA.  A crystal with mass of $\sim 7.5$~mg was aligned with the $c$ axis vertical, and then sealed in an aluminum can with helium exchange gas.  A vertical field cryomagnet was used, providing a base temperature of 1.5~K and a maximum field of $H = 6$~T.  Measurements were performed in phase I at $T$ = 1.5~K and no applied field rotating through 180° in 0.1° steps over 20 h.  Measurements in phase II were made at $T$ = 1.5~K and $H = 3.2$~T by rotating through 180° in 0.1° steps over 20 h. In phase III at $T = 1.5$~K at $H = 4.5$~T the sample was rotated through 90° in 0.1° steps over 10 h.  In phase IV, at $T = 1.5$~K and $H = 5.5$~T the sample was rotated through 90° in 0.1° steps over 10 h.  In phase V, at $T = 4$~K and $H = 5.5$~T the sample was rotated through 90° in 0.1° steps over 10 h.  In phase VI, at $T = 5.05$~K and $H = 5.5$~T the sample was rotated over 90° in 0.1° steps with a total measurement time of 26 h.

%---------------------------------------------------------------------
\begin{figure}[h]
    \includegraphics[width=1.0\textwidth]{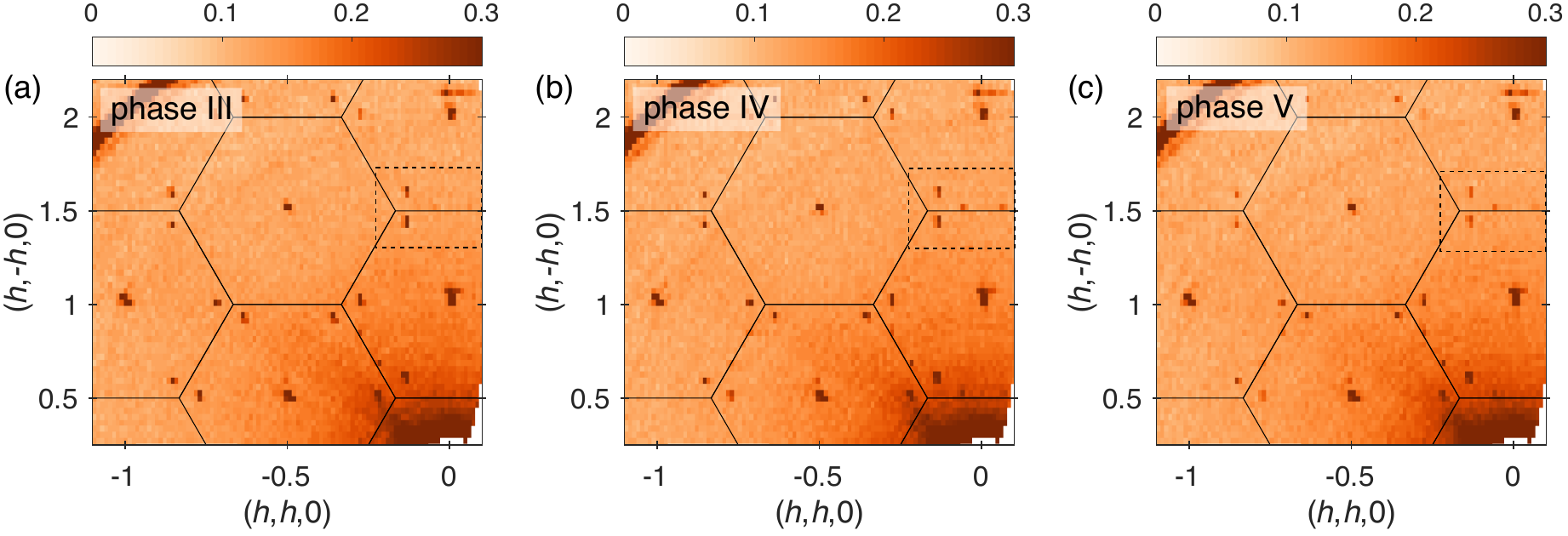}
    \caption{Single crystal neutron diffraction pattern for Cs$_3$Fe$_2$Cl$_9$ in phases (a) III, (b) IV, and (c) V in the ($h$,$k$,0) plane. The experimental conditions are listed in the text. The dashed rectangle emphasizes the area in which the relative intensity of the $\bm{q}_\textrm{III}$ and  $2\bm{q}_\textrm{III}$ reflections are compared in the text.
    \label{fig:wand2}}
\end{figure}
%---------------------------------------------------------------------
%---------------------------------------------------------------------
\begin{figure}[b!]
    \includegraphics[width=0.5\textwidth]{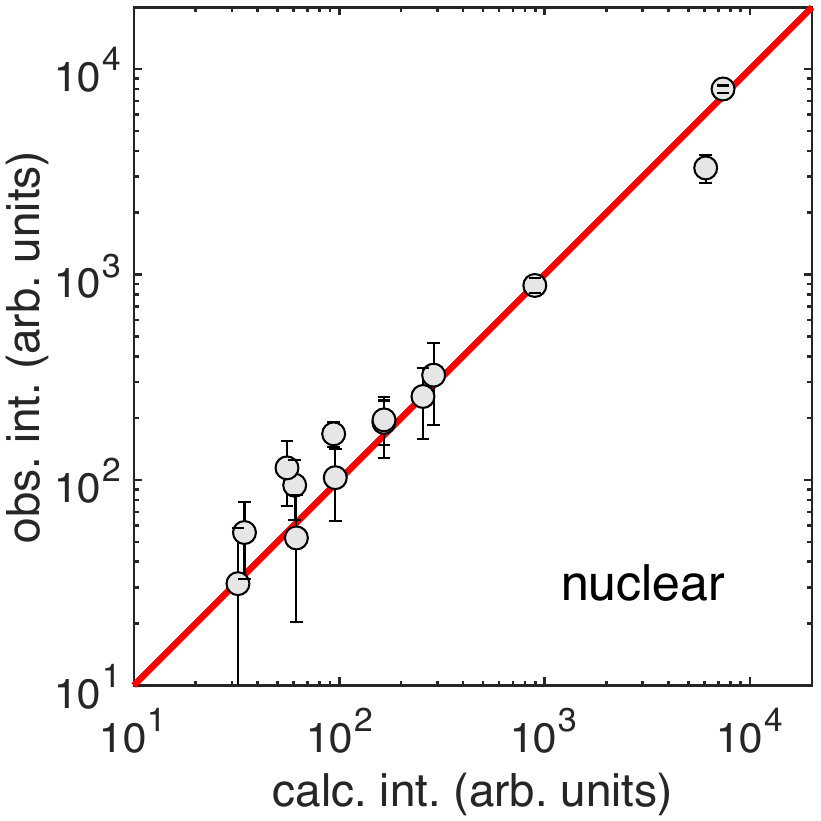}
    \caption{Comparison between the calculated and observed intensity of the nuclear reflections collected on WAND$^2$.
    \label{fig:xtal1}}
\end{figure}
%---------------------------------------------------------------------

Figure~\ref{fig:wand2} compares the neutron diffraction patterns in phases III, IV, and V. As explained in the main text, similar diffraction patterns were observed in these three phases. Especially, their magnetic propagation vector stays at the same $\bm{q}_\textrm{III} = \frac{5}{6}\times(\frac{1}{3}, \frac{1}{3}, 0)$, with weak $2\bm{q}_\textrm{III}$ higher harmonics observed along the Brillouin zone boundary. Despite their similarities, careful comparisons of the diffraction patterns reveal some differences among these three phases. As emphasized by the dashed rectangle in each panel, the intensity of the $2\bm{q}_\textrm{III}$ reflection  in phase III is weaker than that in phase IV, which suggests higher field-induced magnetization in phase IV. As compared in Fig. 4 of the main text for phases IV and V, the $\bm{q}_\textrm{III}$ reflection becomes weaker in phase V while the $2\bm{q}_\textrm{III}$ reflection stays almost unchanged. This evolution is consistent with a transition from a spiral structure in phase IV to a spin density wave (SDW) structure in phase V. For a spiral order like that in phase IV, both the in-plane and out-of-plane components contribute to the intensity of the main $\bm{q}_\textrm{III}$ reflections, while only the out-of-plane component, under renormalization as described in the section of ‘Analytical expressions for the field-induced phases’, contributes to the $2\bm{q}_\textrm{III}$ reflections. For the collinear SDW order in phase V, the in-plane component becomes zero, thus the intensity of the main $\bm{q}_\textrm{III}$ reflections will be strongly reduced compared that in phase IV.

\section{Refinements of the nuclear and magnetic structures}

Figure~\ref{fig:xtal1} compares the experimental intensity of the nuclear Bragg reflections observed on WAND$^2$ to the calculations assuming the same crystal structure as that listed in Table~\ref{tab:refine2}. The goodness-of-fit parameters are $R_{F2}=26.9\%$ and $R_{F2w}=22.4\%$. The agreement between the observed and calculated intensity confirms the good quality of our crystal sample.

%---------------------------------------------------------------------
\begin{figure}[b!]
    \includegraphics[width=0.8\textwidth]{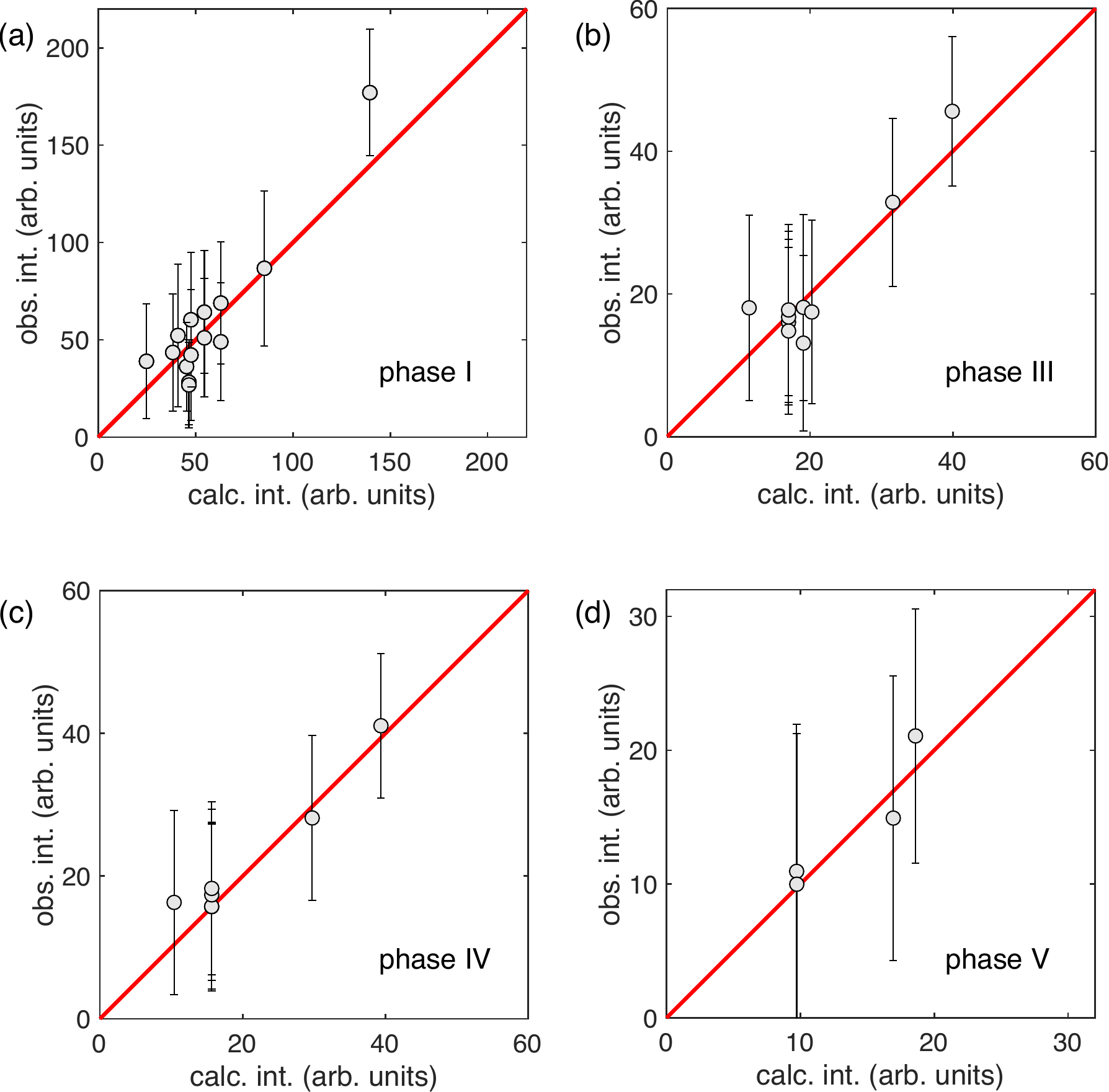}
    \caption{Comparison between the calculated and observed intensity of the magnetic reflections collected on WAND$^2$ for phase I (a), III (b), IV (c), and V (d).
    \label{fig:xtal2}}
\end{figure}
%---------------------------------------------------------------------

Figure~\ref{fig:xtal2} summarizes the fitting results of the magnetic reflections collected on WAND$^2$ in phases I (a), III (b), IV (c), and V (d). The measuring conditions are listed in the main text. In phase VI, only two nonequivalent magnetic reflections were obtained for each magnetic domain, which does not allow a reliable analysis of the magnetic structure. For phase I, the collinear magnetic structure as shown in Fig.~1(a) of the main text was assumed in the fits. The fitted magnitude of the ordered magnetic moment is 2.9(2)~$\mu\rm{_B}$, with goodness-of-fit parameters $R_{F2}=20.3\%$ and $R_{F2w}=24.4\%$. For phases III-V, a helical magnetic structure was assumed based on our classical Monte Carlo simulations. The magnitudes of the in-plane magnetic moment, $M_\parallel$, and out-of-plane magnetic moment, $M_\perp$, were treated as fitting parameters. For phase III, the fitted magnitudes are $M_\parallel = 1.9\pm1.5$ and $M_\perp = 1.9\pm0.4$~$\mu\rm{_B}$, with goodness-of-fit parameters $R_{F2}=13.0\%$ and $R_{F2w}=14.7\%$. For phase IV, the fitted magnitudes are $M_\parallel = 2.1\pm1.6$ and $M_\perp = 1.8\pm0.5$~$\mu\rm{_B}$, with $R_{F2}=9.0\%$ and $R_{F2w}=10.1\%$. For phase V, the fitted magnitudes are $M_\parallel = 1.8\pm4.1$ and $M_\perp = 1.2\pm0.8$~$\mu\rm{_B}$, with $R_{F2}=3.2\%$ and $R_{F2w}=3.7\%$. Considering the standard deviations, the fitted magnitude of the in-plane moment $M_\parallel$ is consistent with our classical Monte Carlo simulations, where phases III and IV exhibit an elongated helical structure, and phase V exhibits a spin density wave structure with only the $S_z$ components.

\section{Temperature evolution of the diffraction pattern}
%---------------------------------------------------------------------

\begin{figure}[t!]
    \includegraphics[width=1.0\textwidth]{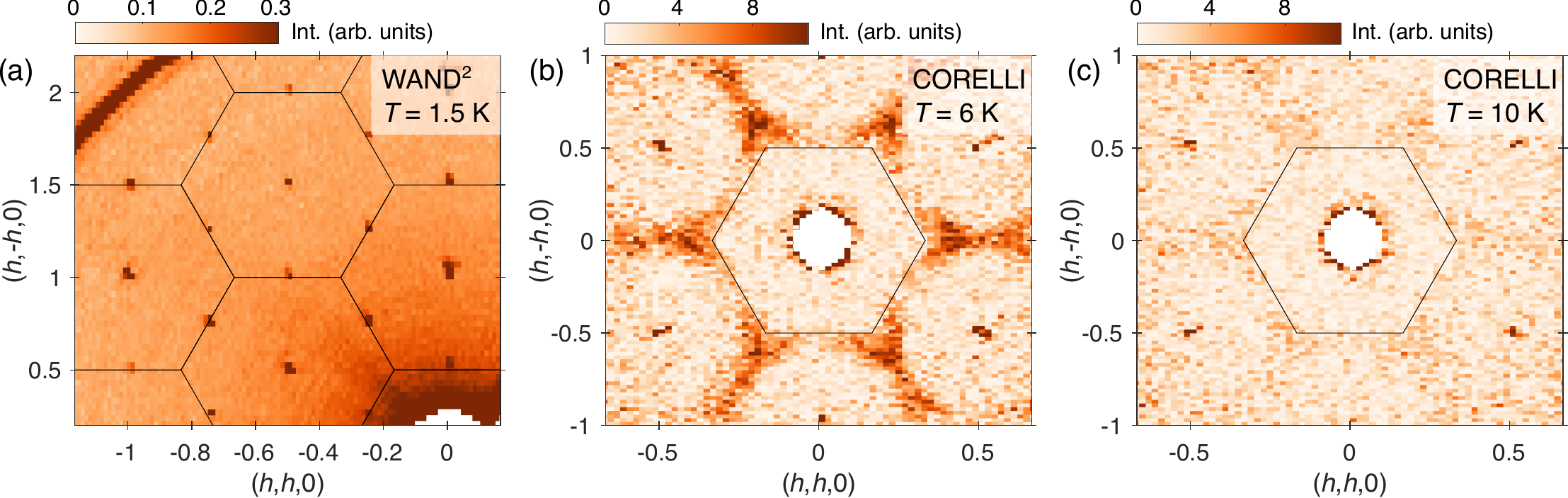}
    \caption{Temperature evolution of the neutron diffraction pattern in the ($h$,$k$,0) plane. The pattern in panel (a) was measured on the WAND$^2$ diffractometer. The patterns in (b) and (c) were measured on the CORELLI diffractometer, and data collected at $T = 20$~K has been subtracted as background.
    \label{fig:tdep}}
\end{figure}

Figure~\ref{fig:tdep} presents the temperature evolution of the neutron diffraction pattern for Cs$_3$Fe$_2$Cl$_9$ in the ($h$,$k$,0) plane. Experiments were measured on the WAND$^2$ ($T = 1.5$~K) and CORELLI ($T=6$ and 10~K) diffractometers using the same piece of crystal. The enhanced diffuse scattering intensity at $T = 6$~K confirms its magnetic origin, which is replaced by magnetic Bragg peaks at $T = 1.5$~K.

%---------------------------------------------------------------------

\section{Modeling the spin dynamics}

To model the magnon dispersion observed in the CNCS experiments, we performed linear spin wave theory (LSWT) calculations using the SpinW software package~\cite{toth_linear_2015s}. The powder-averaged dynamical structure factor, $S(Q, \omega)$, was calculated in steps of $0.05$~\AA$^{-1}$ over a momentum transfer range of $Q = 0.35$ to $2.05$~\AA$^{-1}$ for the spectra collected with $E_i = 3.32$~meV, and over the range of $Q = 0.4$ to $0.95$~\AA$^{-1}$  for the spectra collected with $E_i = 1.0$~meV. The calculated spectrum was convoluted with the theoretical instrumental energy resolution for each $E_i$ configuration plus an additional Gaussian broadening of 0.05~meV that accounts for thermal broadening. To reduce the impacts of neutron absorption, data below and above the energy transfer of $\Delta E = 1.25$~meV in the $E_i = 3.32$~meV spectra were treated separately with a fitted intensity ratio of about 3:2. The optimal exchange parameters were determined through a simultaneous fit for all three datasets including the low- and high-energy sections from the $E_i = 3.32$~meV spectra and the $E_i = 1$~meV spectra, using a weighting factor of 1:3:5, respectively.

%---------------------------------------------------------------------
\begin{figure}[t!]
    \includegraphics[width=0.91\textwidth]{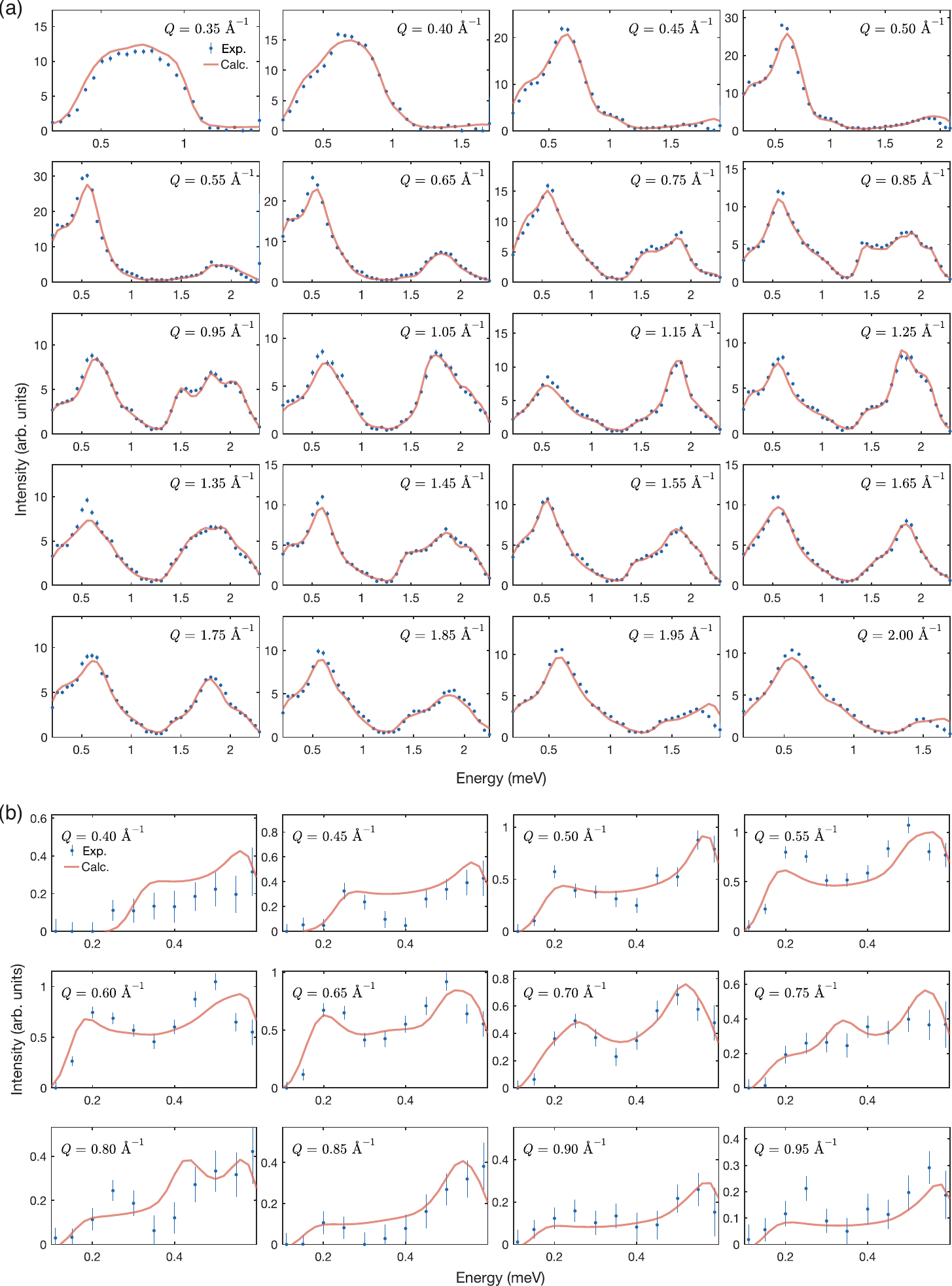}
    \caption{Fit to the INS data using the $J_{1\textrm{-}5}$-$D_z$ model with (a) $E_\mathrm{i} = $ 3~meV data and (b) $E_\mathrm{i} = $ 1~meV data. 
    \label{fig:fit_INS}}
\end{figure}
%---------------------------------------------------------------------

We began the INS fit with an initial $J_{123}$-$D_z$ model, which is the minimum model to realize spiral spin-liquids on the lattice of Cs$_3$Fe$_2$Cl$_9$:
\begin{equation}
    \mathcal{H} = \sum_{n=1,2,3}J_{n}\bm{S}_i\cdot\bm{S}_j + D_z(S_z)^2\textrm{.}
    \label{eqn:model}
\end{equation}
Starting from this minimum model that includes exchange interactions up to the third neighbors, we progressively includes further-neighbor exchange interactions to check whether they can further lower the $\chi^2$ factor. Through this procedure, we found that interactions up to the fourth neighbors, i.e., $J_1$ through $J_4$, are essential for accurately reproducing the INS spectra. The fifth-neighbor interactions, $J_5$, were also included in our model, as it slightly reduces the $\chi^2$ from 20.66 to 20.29, although its fitted strength of 0.0013(1)~meV is already much weaker than the other terms. Further perturbations up to the seventh neighbors yield no improvement for the fit, and their fitted strengths, when included, are all in the orders of $10^{-4}$~meV or even less. We also investigated the impacts of symmetry-allowed anisotropic terms, including the Dzyaloshinskii-Moriya (DM) interactions or off-diagonal exchange terms, but no improvement was observed after introducing these terms in the fitting program. Therefore, in the final $J_{1\textrm{-}5}$-$D_z$ model, we considered only the isotropic exchange interactions up to the fifth neighbors. The fitting results for the INS spectra at representative $Q$ positions are summarized in Fig.~\ref{fig:fit_INS}.

\section{Classical Monte Carlo simulations}
Classical Monte Carlo simulations for the $J_{1\textrm{-}5}$-$D_z$ model were performed using the SpinMC code that implements the single spin flip Metropolis algorithm~\cite{buessen_spinmc_2020}. Unless otherwise specified, a $16\times16\times4$  supercell with 4096 spins was employed in simulations. Observables, including the heat capacity and magnetization, were averaged over $10^5$ of measurement sweeps after $2\times10^5$ thermalization sweeps, where each sweep represents 8192 attempted spin flips at randomly selected sites. An overrelaxation sweep was applied after each Monte Carlo sweep to reduce the autocorrelation. Magnetic structure factors were calculated through fast Fourier transform using the FFTW package~\cite{fftw}. The parallel tempering algorithm was utilized to facilitate thermal equilibrium, which was performed simultaneously over 80 replicas on 80 cores with a geometric series of temperatures between 1 and 10~K. After every 10 Monte Carlo sweeps, a replica exchange was attempted, and the successful exchange rates are above $\sim25\%$ for all the neighboring replicas.

\section{Phase diagram for the $J_{1\textrm{-}5}$-$D_z$ model}

% As discussed in the previous section,  AFM $J_4$ and FM $J_5$ belong to the second group of perturbations and have similar effects on the phase diagram. However, there is one slight difference for these two perturbations: In zero field, AFM $J_4$ and FM $J_5$ raises and lowers $T_N$, respectively. This observation favors FM $J_5$ since the experimental $T_N\sim5.4$~K is lower than the theoretically predicted $T_N\sim5.8$~K for the unperturbed $J_{123}+D_z$ model. 

\begin{figure}[t!]
    \includegraphics[width=1.0\textwidth]{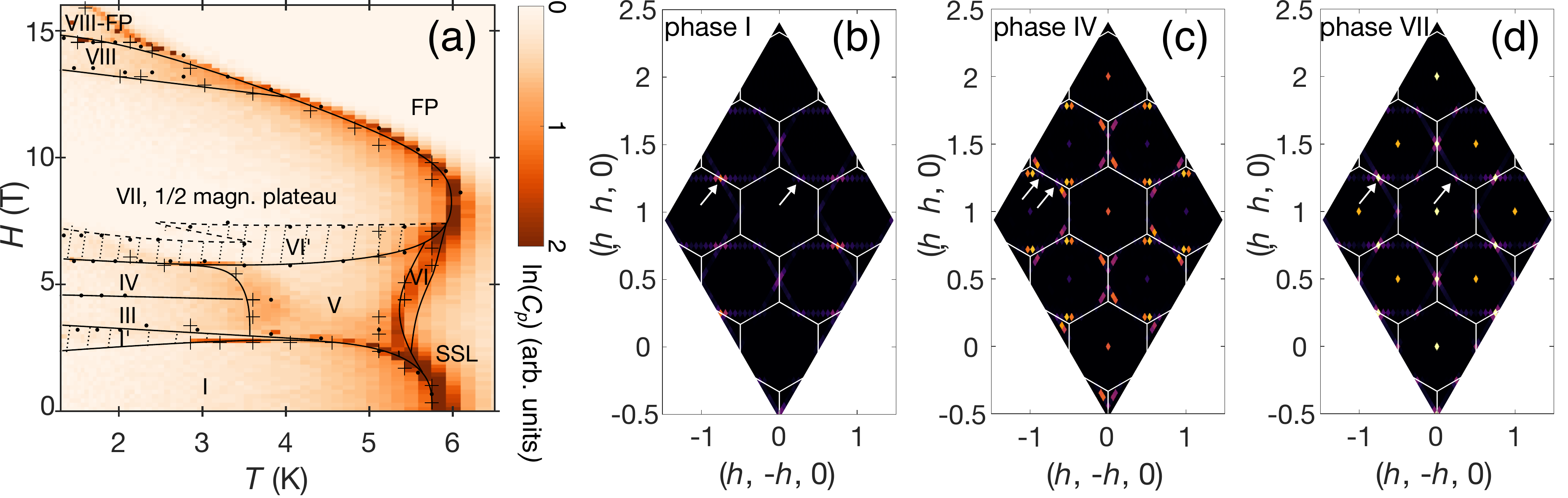}
    \caption{(a) Theoretical phase diagram for the $J_{1\textrm{-}5}$-$D_z$ model calculated by the classical Monte Carlo simulations. Pseudocolor corresponds to the value of heat capacity in arbitrary units (arb. units). (b-d) Magnetic structure factors $\langle \bm{S}\cdot \bm{S} \rangle$ calculated at $T = 1.5$~K in a field of (b) 0~T, (c) 5~T, and (d) 10~T. Calculations of the magnetic structure factor were performed on a $16\times 16\times 4$ supercell with $2\times 10^5$ thermalization sweeps and $1\times10^5$ measurement sweeps. White arrows indicate the positions of weak magnetic Bragg peaks.
    \label{fig:mc_J12345D}}
\end{figure}

\begin{figure}[t!]
    \includegraphics[width=0.4\textwidth]{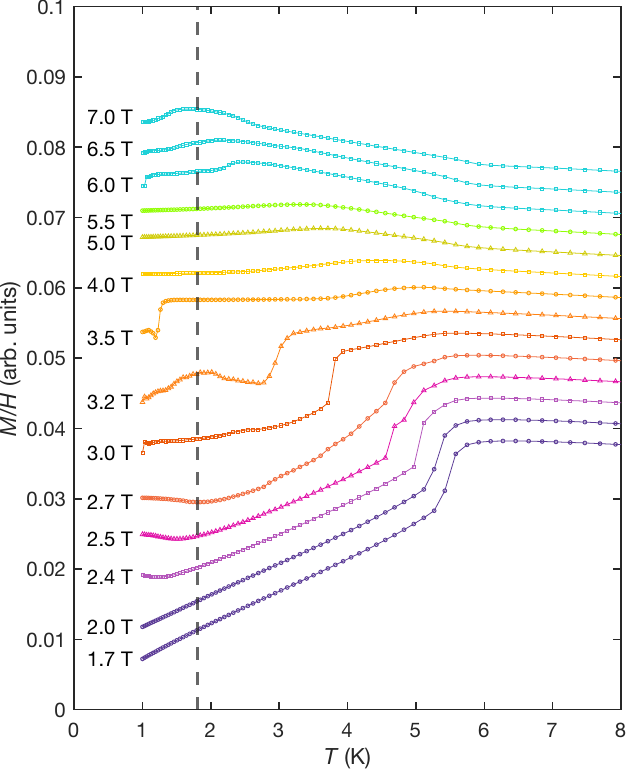}
    \caption{Theoretical magnetization curves calculated by the classical Monte Carlo simulations for the $J_{1\textrm{-}5}$-$D_z$ model. The strength of the applied field for each curve is indicated on the left. Data are successively shifted by 0.003 arb.~units along the $y$ axis for clarity. The dashed line at $T = 1.8$~K marks the lower limit of the measuring temperature in Ref.~\cite{ishii_field_2021s}.
    \label{fig:mc_chiT}}
\end{figure}

\begin{figure}[b!]
    \includegraphics[width=0.95\textwidth]{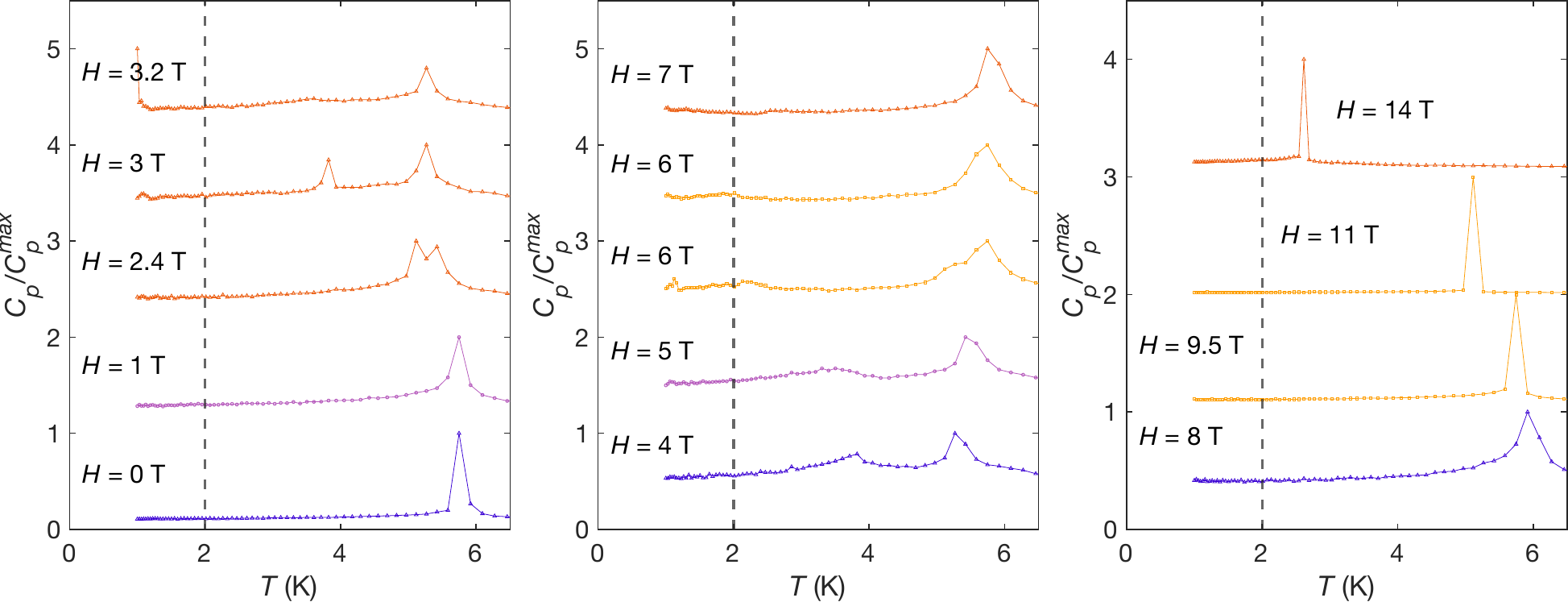}
    \caption{Theoretical heat capacity $C_p$ calculated by the classical Monte Carlo simulations for the $J_{1\textrm{-}5}$-$D_z$ model in the field regime of (a) [0, 3.2], (b) [4, 7], and (c) [8, 14]~T. Data are normalized by the maximal $C_p$ value, $C_p^\textrm{max}$. The strength of the applied field for each curve is indicated on the left. In each panel, data are successively shifted by 1 unit along the $y$ axis for clarity. The dashed line at $T = 2.0$~K marks the lower limit of the measuring temperature in Ref.~\cite{ishii_field_2021s}.
    \label{fig:mc_cp}}
\end{figure}

Figures~\ref{fig:mc_J12345D}-\ref{fig:mc_cp} summarize the simulation results for the $J_{1\textrm{-}5}$-$D_z$ model. By carefully analyzing the field and temperature dependence of the magnetic susceptibility (see Fig.~\ref{fig:mc_chiT}) and specific heat (see Fig.~\ref{fig:mc_cp}), we obtain the phase diagram shown in Fig.~\ref{fig:mc_J12345D}(a). The typical magnetic structure factors shown in Figs.~\ref{fig:mc_J12345D}(b-d) reproduce the $\bm{q}_\textrm{I} = (\frac{1}{2}, 0, 0)$ in phase I and VII at $H=$~0 and 10~T, respectively, and $\bm{q}_\textrm{III} = \frac{5}{6}\times(\frac{1}{3}, \frac{1}{3}, 0)$ in phase IV at $H = 5$~T.

%---------------------------------------------------------------------
%---------------------------------------------------------------------

\begin{figure}[htb]
    \includegraphics[width=1.0\textwidth]{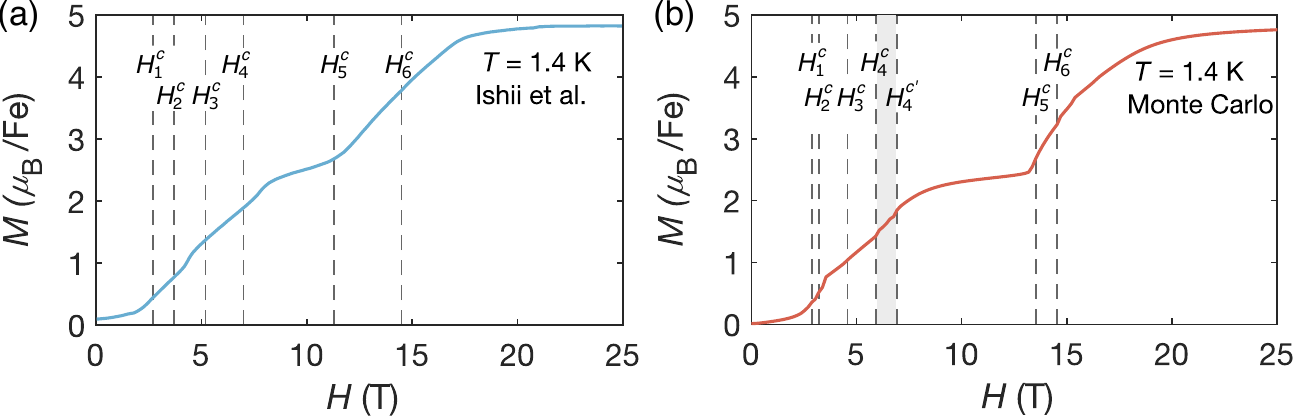}
    \caption{Experimental (a) and theoretical (b) magnetization curves.  The experimental data, together with the phase boundaries indicated by the dashed lines, are adapted from Ref.~\cite{ishii_field_2021s}. The theoretical data were obtained from classical Monte Carlo simulations using the  $J_{1\textrm{-}5}$-$D_z$ model.
    \label{fig:MH}}
\end{figure}

Figure~\ref{fig:MH} compares the experimental and theoretical magnetization curves calculated by classical Monte Carlo simulations for the fitted $J_{1\textrm{-}5}$-$D_z$ model. The identified critical fields are largely consistent with those in the phase diagram in Fig.~\ref{fig:mc_J12345D}(a). The existence of the 1/2 magnetization plateau phase is well reproduced in our simulations.

%---------------------------------------------------------------------
%---------------------------------------------------------------------
%---------------------------------------------------------------------
%---------------------------------------------------------------------
%---------------------------------------------------------------------
%---------------------------------------------------------------------
%---------------------------------------------------------------------
\begin{figure}[h!]
    \includegraphics[width=1.0\textwidth]{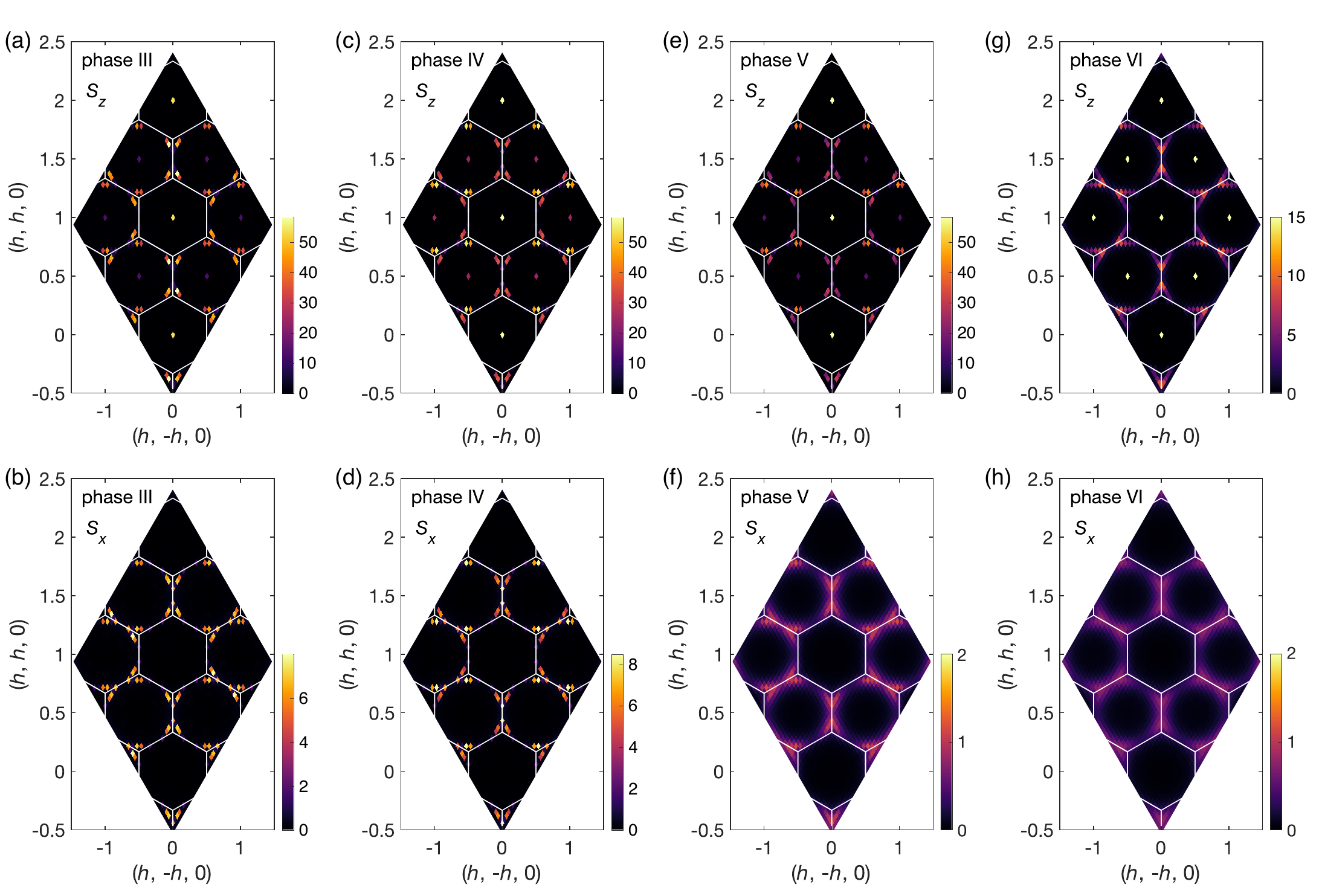}
    \caption{Theoretical magnetic structure factors (a) $\langle S_zS_z \rangle$ and (b) $\langle S_xS_x \rangle$ in phase III for the $J_{1\textrm{-}5}$-$D_z$ model. Calculations were performed on a $16\times 16\times4$ supercell using the classical Monte Carlo simulations. The temperature and magnetic field are set to 2.33~K and 4~T, respectively. Calculations were performed over $1\times10^5$ measuring sweeps after $2\times10^5$ thermalization sweeps. (c-h) Similar magnetic structure factors in phases IV, V, and VI. The calculation parameters are $T=2.33$~K and $H = 5$~T in phase IV, $T=4.17$~K and $H = 4.0$~T in phase V, and $T=5.58$~K and $H = 5.0$~T in phase VI.
    \label{fig:mc_strFac}}
\end{figure}
%---------------------------------------------------------------------

\subsection{Spiral-type orders in phases III and IV}

Comparisons of the direction-resolved magnetic structure factors and real-space spin configurations unveil the magnetic structures in each of the field-induced phases. Figure~\ref{fig:mc_strFac} compares the $\langle S_zS_z\rangle$ and $\langle S_xS_x\rangle$ components of the magnetic structure factors calculated for phases III, IV, V, and VI through the classical Monte Carlo simulations. In phases III and IV, both the $\langle S_zS_z\rangle$ and $\langle S_xS_x\rangle$ components exhibit long-range order, suggesting a spiral-type magnetic order that involves both in-plane and out-of-plane spin components. The structure factor along the $z$ direction is about 10 times higher than that along the $x$ direction, which indicates that the spiral order in phases III and IV is elongated along the $c$ axis due to the uniaxial SIA. Compared to that in phase III, the relatively stronger $2\bm{q}_\textrm{III}$ reflections in phase IV suggest stronger squaring-up effects in a higher magnetic field. This field dependence of the $2\bm{q}_\textrm{III}$ reflections is consistent with the experimental observations shown in Fig.~\ref{fig:wand2}.

\subsection{SDW orders in phases V and VI}

As shown in Fig.~\ref{fig:mc_strFac}(f) and (h), the $\langle S_xS_x\rangle$ component of the magnetic structure factors in phases V and VI exhibits diffuse patterns, while sharp Bragg peaks are observed in the $\langle S_zS_z\rangle$ component. This observation indicates that magnetic moments in phases V and VI are ordered only along the $c$ axis, leading to sinusoidally modulated SDW orders in these two phases. We also note that comparison of the $\langle S_zS_z\rangle$ component in phases III (Fig.~\ref*{fig:mc_strFac}(a)), IV (Fig.~\ref*{fig:mc_strFac}(c)), and V (Fig.~\ref*{fig:mc_strFac}(e)) reveals a similar magnetic propagation vector $\bm{q}_\textrm{III}$ and its high harmonics $2\bm{q}_\textrm{III}$, which reproduces the experimental observations shown in Fig.~\ref{fig:wand2}.

\subsection{Magnetic orders in phases VII and VIII}

\begin{figure}[htb]
    \includegraphics[width=0.95\textwidth]{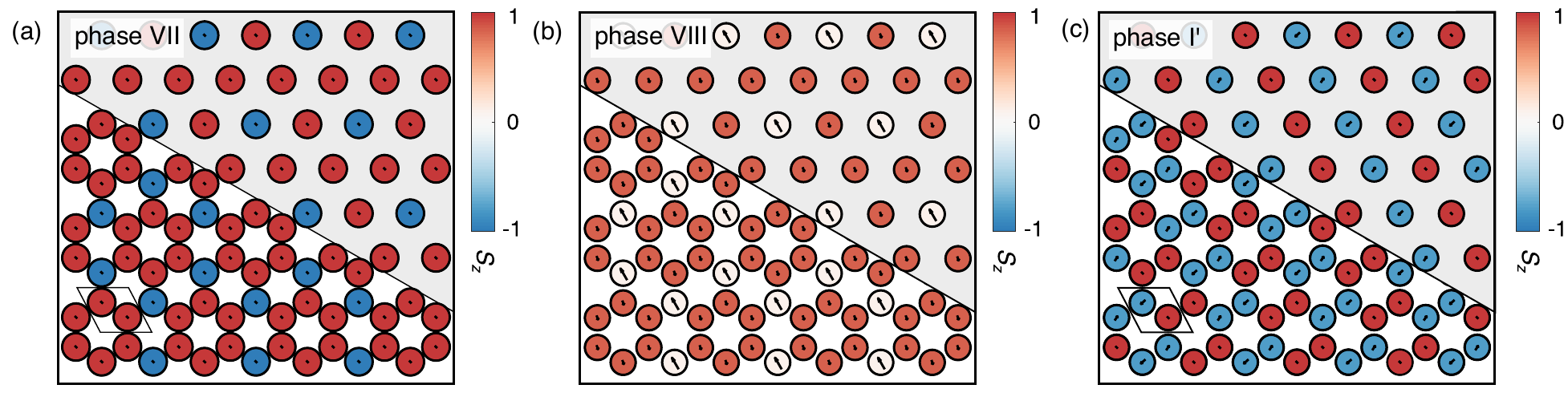}
    \caption{Magnetic orders in phases (a) VII, (b) VIII, and (c) I' viewed along the $c$ axis as determined from classical Monte Carlo simulations for the $J_{1\textrm{-}5}$-$D_z$ model. The bottom left (top right) part depicts the spin configuration for two sublattices (one sublattice).
    \label{fig:plateau}}
\end{figure}

A few possible magnetic orders have been proposed for the 1/2-magnetization plateau phase~\cite{ishii_field_2021s}. Through classical Monte Carlo simulations, the order in the 1/2-magnetization plateau phase of the $J_{1\textrm{-}5}$-$D_z$ model is presented in Fig.~\ref{fig:plateau}(a). In the same figure, the predicted magnetic order in phase VIII of the $J_{1\textrm{-}5}$-$D_z$ model is presented in Fig.~\ref{fig:plateau}(b). Further neutron diffraction experiments in fields above 6~T will be required to verify these magnetic orders.

%---------------------------------------------------------------------
%---------------------------------------------------------------------
\subsection{Additional intermediate phases}
As compared in Fig.~4 of the main text, the calculated phase diagram contains two intermediate phases that are possibly not observed in the experiment. In the regime of the experimental phase II, the calculated structure is a double-$\mathbf{q}$ phase that is denoted as I', where the ordering vectors are the same as those in phase I. The simulated magnetic order in this phase is presented in Fig.~\ref{fig:plateau}(c). Close to the lower phase boundary of the 1/2-plateau phase VII, another transitional phase, designated VI', is identified in the calculated phase diagram. The magnetic order in phase VI' is largely the same as that of phase VI, while the $\langle S_xS_x \rangle$ and $\langle S_yS_y \rangle$ components is enhanced in phase VI' due to its relatively low temperatures.

In our classical Monte Carlo simulations, we did not manage to remove these additional phases, especially phase VI’, by introducing further perturbations to the $J_{1\textrm{-}5}$-$D_z$ model. Therefore, we propose that the absence of this transitional phase in experiments might be due to quantum fluctuation that is not considered in our classical Monte Carlo calculations. On the antiferromagnetic triangular lattice model, quantum fluctuation is known to stabilizes a 1/3 magnetization plateau with a collinear up-up-down magnetic order even in the case of $S = 5/2$~\cite{coletta_semiclassical_2016}. Similarly, in Cs$_3$Fe$_2$Cl$_9$, quantum fluctuation may enhance the stability of the 1/2 magnetization plateau phase and thus remove the transitional VI’ phase obtained in classical Monte Carlo simulations.

\section{Analytical expressions for the field-induced phases}

To further verify the magnetic structures of the field-induced phases, we calculate the magnetic structure factors for the following ansatz on a honeycomb lattice to describe the spiral-type orders in phases III and IV (Eqn.~(\ref{eqn:ansatz1})) and the SDW orders in phases V and VI (Eqn.~(\ref{eqn:ansatz2})):
\begin{equation}
    \bm{M}(\bm{r}) = \left\{  \begin{array}{l} M_\perp\cos(\bm{q}\cdot\bm{r})\bm{n}_1 +  M_\parallel\sin(\bm{q}\cdot\bm{r})\bm{n}_2 + M_z\bm{n}_1  \textrm{,\qquad\qquad\ \ if}\ \bm{r} \in \{r_1\} \\
    M_\perp\cos(\bm{q}\cdot\bm{r}+\phi)\bm{n}_1 +  M_\parallel\sin(\bm{q}\cdot\bm{r}+\phi)\bm{n}_2 + M_z\bm{n}_1 \textrm{,\quad if}\ \bm{r} \in \{r_2\} \end{array} \right.
    \label{eqn:ansatz1}
\end{equation}

\begin{equation}
    \bm{M}(\bm{r}) = \left\{  \begin{array}{l} M_\perp\cos(\bm{q}\cdot\bm{r})\bm{n}_1 + M_z\bm{n}_1  \textrm{,\qquad\ \ if}\ \bm{r} \in \{r_1\} \\
    M_\perp\cos(\bm{q}\cdot\bm{r}+\phi)\bm{n}_1 +  M_z\bm{n}_1 \textrm{,\quad if}\ \bm{r} \in \{r_2\} \end{array} \right.
    \label{eqn:ansatz2}
\end{equation}

%---------------------------------------------------------------------
\begin{figure}[htb]
    \includegraphics[width=0.75\textwidth]{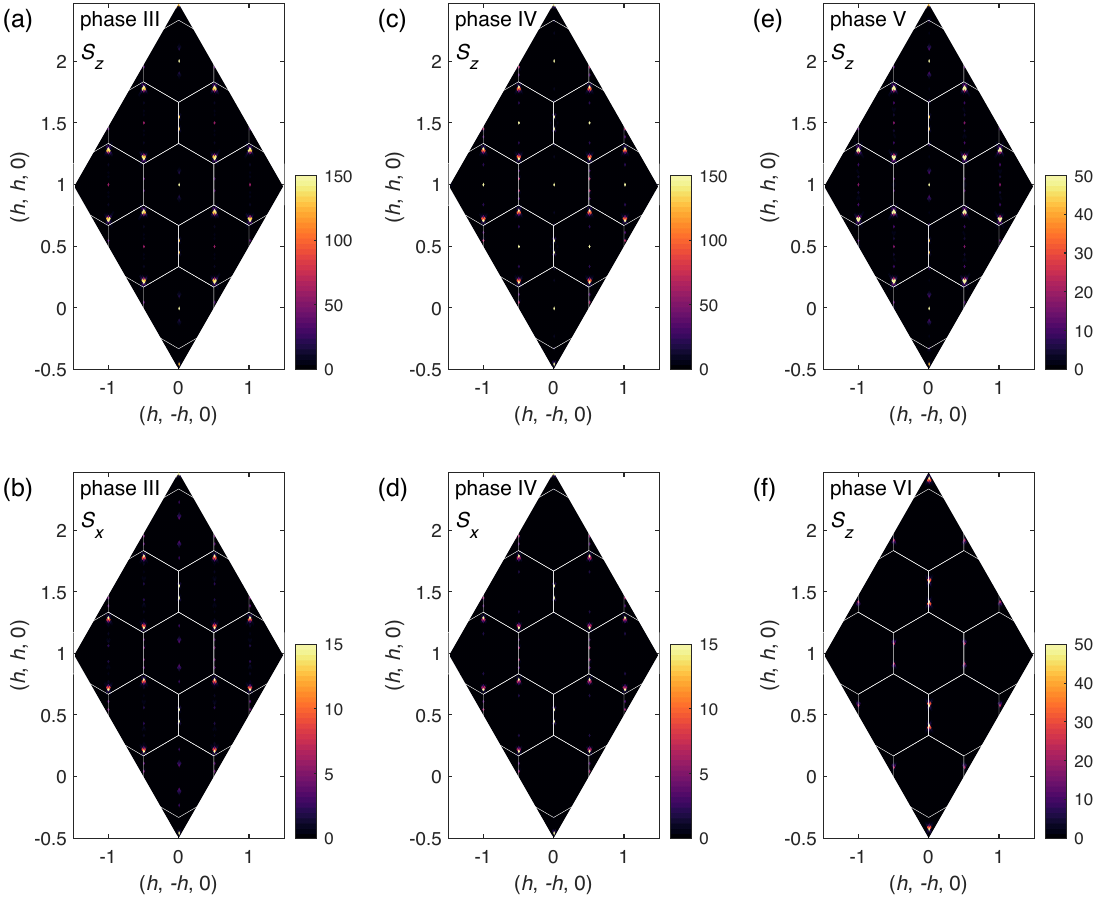}
    \caption{(a) Theoretical magnetic structure factors (a) $\langle S_zS_z \rangle$ and (b) $\langle S_xS_x \rangle$ using the ansatz in Eqn.~(\ref{eqn:ansatz1}) for phase III.  Calculations were performed for a two-dimensional $30\times30$ honeycomb superlattice through the Fourier transform. (c-d) Magnetic structure factors using the ansatz in Eqn.~(\ref{eqn:ansatz1}) for phases IV. (e) Magnetic structure factor $\langle S_zS_z \rangle$ using the ansatz in Eqn.~(\ref{eqn:ansatz2}) for phase V. (d) Magnetic structure factor $\langle S_zS_z \rangle$ using the ansatz in Eqn.~(\ref{eqn:ansatz2}) for phase VI.
    \label{fig:ansatz}}
\end{figure}
%---------------------------------------------------------------------

In these expressions, $M_\perp$ and $M_\parallel$ are the ordered moments perpendicular to and parallel with the $ab$ plane, respectively. $M_z$ is the field induced moment along the $c$ axis. $\bm{n}_1$ is a unit vector along the $c$ axis,  $\bm{n}_2$ is defined in a way that the vectors $\bm{q}$, $\bm{n}_2$, and $\bm{n}_1$ form a cartesian coordinate system. An additional phase factor $\phi$ is introduced for spins on the second sublattice ($\bm{r} \in \{r_2\}$).

Figure~\ref{fig:ansatz} summarizes the magnetic structure factors for each ansatz. $\bm{M}(\bm{r})$ in real space was first calculated on a $30\times30$ superlattice, renormalized to equal moment size if $M_z \neq 0$, and then Fourier transferred to reciprocal space. The parameters are $\bm{q}= \bm{q}_\textrm{III} = \frac{5}{6}\times(\frac{1}{3}, \frac{1}{3})$, $M_\perp=0.9$, $M_\parallel = 0.1$, $\phi=\pi$, and $M_z = 0.38$ (0.58) for phase III (IV). The parameters for phase V are $\bm{q}=\bm{q_\textrm{III}} = \frac{5}{6}\times(\frac{1}{3}, \frac{1}{3})$, $M_\perp=1$, $M_z = 0.48$, and $\phi = \pi$. The parameters for phase VI are $\bm{q}_\textrm{VI} = \frac{11}{9}\times(\frac{1}{3}, \frac{1}{3}, 0)$, $M_\perp=0.5$, $M_z = 0$, and $\phi = 0$. By comparing the magnetic structure factors in Fig.~\ref{fig:ansatz} with those in Fig.~\ref{fig:mc_strFac}, Fig.~\ref{fig:wand2}, and Fig.~4 in the main text, it is confirmed that the double peaks along over the Brillouin zone boundaries are due to high harmonics $2\bm{q}_\textrm{III}$ reflections induced by magnetic field, thus corroborating the proposed spiral-type and SDW orders in the field-induced phases.

%---------------------------------------------------------------------
\begin{figure}[t!]
    \includegraphics[width=1\textwidth]{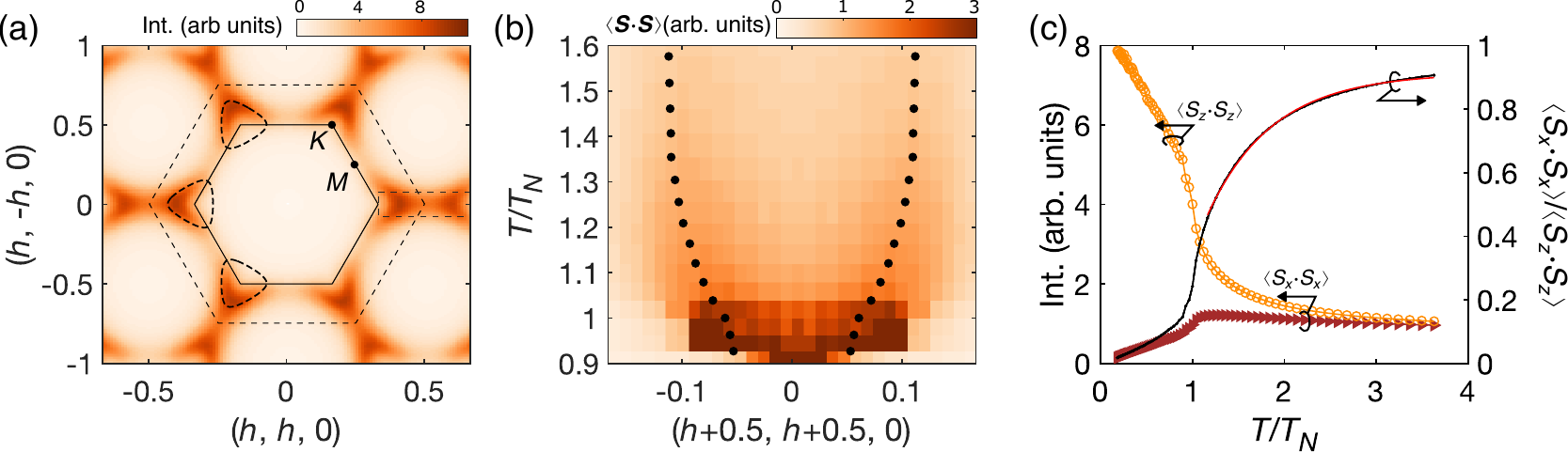}
    \caption{(a) Calculated diffuse scattering pattern in the ($h,~k,~0$) plane using the fitted $J_{123}$-$D_z$ model at $T = 6$~K. (b) Temperature evolution of the calculated spin correlation function $\langle \bm{S}\cdot \bm{S}\rangle$ along ($h$, $h$, 0) around (0.5, 0.5, 0). The integration area is outlined in panel (a) by a dashed rectangle. Black dots indicate the peak positions at the corresponding temperatures through Gaussian fits. (c) Temperature evolution of the calculated spin correlations $\langle S_x\cdot S_x\rangle$ (red triangles), $\langle S_z\cdot S_z\rangle$ (yellow circles), and their ratio $\langle S_x\cdot S_x\rangle / \langle S_z\cdot S_z\rangle$ (black dots). The integration area is outlined in panel (a) by a dashed-line hexagon. Red line over the $\langle S_x\cdot S_x\rangle / \langle S_z\cdot S_z\rangle$ data is the fitting curve to the Arrhenius law $A_1\exp(-\frac{T-T_N}{\Delta_\textrm{SIA}})+A_2$ with fitted parameters of  $A_1 = -0.561(4)$, $\Delta_\textrm{SIA}=3.93(7)$~K, and $A_2=0.913(2)$.
    \label{fig:ising}}
\end{figure}
%---------------------------------------------------------------------

\section{Effects of the uniaxial anisotropy on the SSL}
Classical Monte Carlo simulations for the fitted $J_{1\textrm{-}5}$-$D_z$ model were performed to study the effects of the SIA on the SSL in Cs$_3$Fe$_2$Cl$_9$. As a reference, the diffuse scattering pattern at $T=6$~K calculated by the SCGA method is shown in Fig.~\ref{fig:ising}. Through classical Monte Carlo simulations that incorporate critical scattering near $T_N$, Fig.~\ref{fig:ising}(b) summarizes the temperature evolution of the calculated scattering intensity along ($h+\frac{1}{2}$, $h+\frac{1}{2}$, 0), where the integration range is outlined by dashed rectangle in Fig.~\ref{fig:ising}(b). For the fitted $J_{1\textrm{-}5}$-$D_z$ model, $T_N$ is found to be $\sim5.75$~K. The separated peaks at $T/T_N>1$ correspond to thermally stabilized spiral correlations in the SSL state, while their gradually reduced separation with decreasing $T$ reveals the second-order character of the transition into the LRO. Figure~\ref{fig:ising}(c) compares the in-plane ($x$) and out-of-plane ($z$) components of the calculated spin correlations as a function of $T$. Due to the existence of uniaxial SIA, the evolution of $\langle S_x\cdot S_x\rangle$ is nonmonotonic, and the ratio between the in-plane and out-of-plane components exhibits an Arrhenius-like behavior with an activation gap of  $\Delta_\textrm{SIA} = 3.93(7)$~K in a wide temperature regime. The value of the activation energy is close to the estimation from the single ion anisotropic energy of $D_zS(S+1) \sim 3.27$~K that is required to flip a single spin from the easy axis along $c$ to the $ab$ plane.

\end{document}